\documentclass[structabstract]{aa}  

\usepackage{psfig,hyperref,amssymb,subfigure}
\usepackage[usenames,dvipsnames]{xcolor}
\usepackage{natbib}
\usepackage{graphicx}
\usepackage{amsmath}
\newcommand\beq{\begin{equation}}
\newcommand\eeq{\end{equation}}

\def\msun{\,{\rm M_\odot}}

\def\gsim{ \lower .75ex \hbox{$\sim$} \llap{\raise .27ex \hbox{$>$}} }
\def\lsim{ \lower .75ex\hbox{$\sim$} \llap{\raise .27ex \hbox{$<$}} }

\usepackage{txfonts}
\begin{document}

\title{Evolution of binary black holes in self gravitating discs}

   \subtitle{Dissecting the torques}

   \author{C. Roedig\inst{1}\fnmsep\thanks{E-mail: croedig@aei.mpg.de}
          \and
           A. Sesana\inst{1}
          \and
           M. Dotti \inst{2}
          \and
           J. Cuadra \inst{3}
          \and
           P. Amaro-Seoane\inst{1,4}
          \and
           F. Haardt\inst{5}
          }

   \institute{ Max-Planck-Institut f{\"u}r Gravitationsphysik, Albert Einstein Institut, Am M\"uhlenberg 1, 14476 Golm, Germany \\
               \and
           Universit\`a di Milano Bicocca, Dipartimento di Fisica, G. Occhialini, Piazza della Scienza 3, I-20126, Milano, Italy\\
              \and
           Departamento de Astronom\'ia y Astrof\'isica, Pontificia Universidad Cat\'olica de Chile, 7820436 Macul, Santiago, Chile\\
              \and
             Institut de Ci{\`e}ncies de l'Espai (CSIC-IEEC), Campus UAB, Torre C-5, parells, $2^{\rm na}$ planta, ES-08193, Bellaterra, Barcelona, Spain\\
             \and
             Dipartimento di Scienza e Alta Tecnologia, Universit\'a dell'Insubria, via Valleggio 11, 22100 Como, Italy
             }

   \date{Received 11-07-2012; accepted 13-08-2012}

 
  \abstract
   {Massive black hole binaries, formed in galaxy mergers, are
expected to evolve in dense circumbinary discs. Understanding of
the disc-binary coupled dynamics is vital to assess both the final
fate of the system and the potential observable features that may be
tested against observations.}
   {Aimed at
  understanding the physical roots of the secular evolution of the
  binary, we study the interplay between gas accretion and gravity
  torques in changing the binary elements (semi-major axis and
  eccentricity) and its total angular momentum budget. We pay special
  attention to the gravity torques, by analysing their physical origin
  and location within the disc.
   }
   { We analyse three-dimensional smoothed particle hydrodynamics
  simulations of the evolution of initially quasi-circular massive
  black hole binaries (BHBs) residing in the central hollow (\textit{cavity}) of massive
  self-gravitating circumbinary discs.  We perform a set of
  simulations adopting different thermodynamics for the gas within the
  cavity and for the 'numerical size' of the black holes.
   }
   {We show that (i) the BHB eccentricity growth found in our previous work
   is a general result, independent of the accretion and the adopted 
  thermodynamics; 
  (ii) the semi-major axis decay depends not only on the gravity 
  torques but also on their subtle interplay with the disc-binary 
  angular momentum transfer due to accretion; 
  (iii) the spectral structure of the gravity torques is predominately 
  caused by disc edge overdensities and spiral arms developing in 
  the body of the disc and, in general, does not reflect directly the 
  period of the binary; 
  (iv) the net gravity torque changes sign 
  across the BHB corotation radius (positive inside vs negative outside)
  We quantify the relative importance 
  of the two, which appear to depend on the thermodynamical properties of 
  the instreaming gas, and which is crucial in assessing the disc--binary 
  angular momentum transfer; 
  (v) the net torque manifests as a purely kinematic (non-resonant) effect as it stems 
  from the low density cavity, where the material flows in and out in highly 
  eccentric orbits.
   }
   {
  Both accretion onto the black holes and the interaction with gas 
  streams inside the cavity must 
  be taken into account to assess the fate of the binary. Moreover, 
  the total torque exerted by the disc affects the binary angular
  momentum by changing all the elements (mass, mass ratio, eccentricity, 
  semimajor axis) of the black hole pair. Commonly used prescriptions equating
  tidal torque to semi-major axis shrinking might therefore be poor approximations
  for real astrophysical systems.
  }

\keywords{
Black hole physics -- Accretion, accretion discs -- Numerical -- Hydrodynamics
}

\maketitle
\section{Introduction}

In the currently favoured hierarchical framework of structure formation
\citep{White78}, galaxies evolve through a complex sequence of merger
and accretion events, and the existence of massive black holes (BHs)
at their centres is nowadays a well established observational fact
\citep[see][and references therein]{gultekin09}.  By combining
together these two pieces of information, the formation of a large
number of massive black hole binaries (BHBs), following galaxy
mergers throughout cosmic history, is a natural consequence of the
structure formation process \citep{begelman80}. Although this is
corroborated by several observed quasar pairs at a $\sim$ 100kpc
projected separation \citep{Hennawi06,Myers07,Myers08,Foreman09,Shen11}, and by few
$\lsim$ kpc dual accreting BHs embedded in the same galaxy
\citep[e.g.][]{Komossa:2002tn,Fabbiano11}, identification of gravitationally
bound BHBs in galaxy centres remains elusive \citep[for an up-to-date
review on the candidates see][and references therein]{Dotti12}.

Even though observationally there is little evidence for their
existence, much theoretical work has focused lately on Keplerian
massive BHBs residing in galactic nuclei. One reason is that 
a deep understanding of the interplay between BHBs and their dense (stellar and
gaseous) environment is required to predict robust signatures that may
allow their identification. Additionally, it is still unclear how Nature
bridges the gap between the two theoretically well understood stages 
of BHB evolution: (i) the dynamical friction driven stage, when the two BHs
spiral in toward the centre of the merger remnant down to pc
separations and (ii) the final inspiral driven by gravitational waves
(GWs), which become efficient when the two BHs are at a separation
$\lsim10^{-2}$ pc.  Both dense stellar and gaseous environments have
been shown to be effective in extracting the binary energy and angular
momentum \citep[see, e.g.,][]{Escala2005,Dotti07,Jorge09,
  Khan11,Preto11}, likely driving the system to final coalescence
\citep[an extensive discussion on the fate of sub-parsec BHBs can be
  found in][]{Dotti12}.  Scenarios involving cold gas are particular
appealing not only because they might produce distinctive
observational signatures, but also because cold gas dominates the
baryonic content in most galaxies at redshifts higher than one,
providing a natural reservoir of energy and angular momentum to drive
the BHB towards coalescence.

Numerical simulations of wet galaxy mergers \citep[e.g.,][]{mihos96,Mayer07}
have led to the following picture for the post merger evolution.  Cold
gas is funnelled to the central $\approx 100$ pc by gravitational
instabilities, where it forms a puffed, rotationally supported
circumnuclear disc. Disc-like structures are actually observed in
ULIRGs which are thought to be gas rich post-merger star-forming
galaxies \citep[e.g.,][]{Sanders96,Downes98,Davies04a,Davies04b,Greve09}. 
In the models, the two nuclear BHs efficiently spiral to sub-pc scales owing to
dynamical friction against the massive circumnuclear disc
\citep{Dotti07}, eventually opening a cavity (or hollow) in the gas
distribution \citep{GoldreichTremaine80}. The subsequent evolution of
the system is determined by the efficiency of energy and angular
momentum transfer between the BHB and its outer circumbinary disc.

The investigation of coupled disc--binary systems has a long-standing
tradition in the context of planetary dynamics
\citep{GoldreichTremaine80,Lin86,Ward97,Bryden99,Lubow99,Nelson00},
where the focus usually lies on the extreme mass ratio situation,
i.e., a star surrounded by a circumstellar disc, with a planetary
companion embedded in it. The comparable mass case has also been
extensively investigated in the context of binary star formation
\citep{Artymowicz1994, Bate97,Gunther02,Gunther04}, where a boost of
activity has been triggered by imaging of nearby young binary stars
embedded in hollow circumbinary discs \citep{Dutrey94}. More recently,
the techniques adopted in these fields have been applied to the BHB
case. In the last decade, several investigations were devoted to the
study of comparable mass BHBs evolving in circumbinary discs,
exploiting a variety of analytical and numerical techniques
\citep{Ivanov99,Armitage:2002,Armitage:2005,hayasaki07,haya08,MacFadyen2008,hayasaki09,Jorge09,
  Lodato2009,nixon11a,Roedig2011,Shi11}. However only few of them
focused on the details of the dynamical disc--binary
interplay. \cite{MacFadyen2008} (hereinafter MM08) made use of
two-dimensional grid-based hydrodynamical simulation to study a BHB
embedded in a thin $\alpha$-disk \citep{ss73}. They showed that the
gas flowing through the cavity increases the energy and angular momentum
transfer between the disc and the binary. \cite{Shi11} (hereinafter
S11) confirmed that result using
full 3D magnetohydrodynamics (MHD) simulations. However, in both
studies the BHB was on a fixed circular orbit, the central region
of the disc (i.e.~within twice the binary separation) was excised from the
computational domain, and the disc self-gravity was neglected.

We employ full 3D smoothed particle hydrodynamical (SPH) simulations
to study the interaction between BHBs and their surrounding discs. Our
goal is to give a detailed description of the coupled disc--binary
dynamics, paying particular attention to the competing effects of
disc--binary gravitational torques and gas accretion onto the BHs in the
evolution of the binary angular momentum budget. We simultaneously
evolve the disc and the two BHs in a self-consistent way. This enables
us to separately investigate the effect of gravitational torques
coming from different disc regions and to directly link different
physical mechanisms (gravitational torques and accretion) to the
evolution of individual quantities describing the binary (mass, mass
ratio, semi-major axis and eccentricity). This is different than 
previous studies (MM08, S11), where the binary was modelled as a fixed forcing
quadrupolar potential and the central region was excised.
To make sure that our results are not spuriously dependent
on particular choices of sensitive parameters,
 we performed different runs varying physical and numerical prescriptions
that may play a relevant role in the disc--binary energy and angular
momentum exchanges. Specifically, we checked the possible
dependence of our results on the 'numerical size' of the two BHs
(i.e., their sink radii, see next Section), and on the adopted
\textit{\textbf{e}quation \textbf{o}f \textbf{s}tate} (EoS) 
for the gas within the central cavity.

The paper is structured as follows: we first describe the numerical
setup and initial conditions in Section~\ref{sec:sims}, then we show
the importance of both accretion and gravitational torques on the
binary evolution in Section~\ref{sec:evolution}. We study in detail
the origin and strength of the mutual disc--binary gravitational
torques in Section~\ref{sec:torx} highlighting the influence of the
thermodynamics prescription. In Section~\ref{sec:spin} we briefly
discuss the temporal behaviour and the spatial geometry of the
accretion, speculating on the evolution of the binary mass ratio and
BH individual spins. Finally, we compare our results to previous work
in Section~\ref{sec:comparison} and discuss the implications of our
findings and give our conclusions in Section~\ref{sec:conclusion}.

\section{Simulations}
\label{sec:sims}
\begin{figure}
 \includegraphics[width=\linewidth]{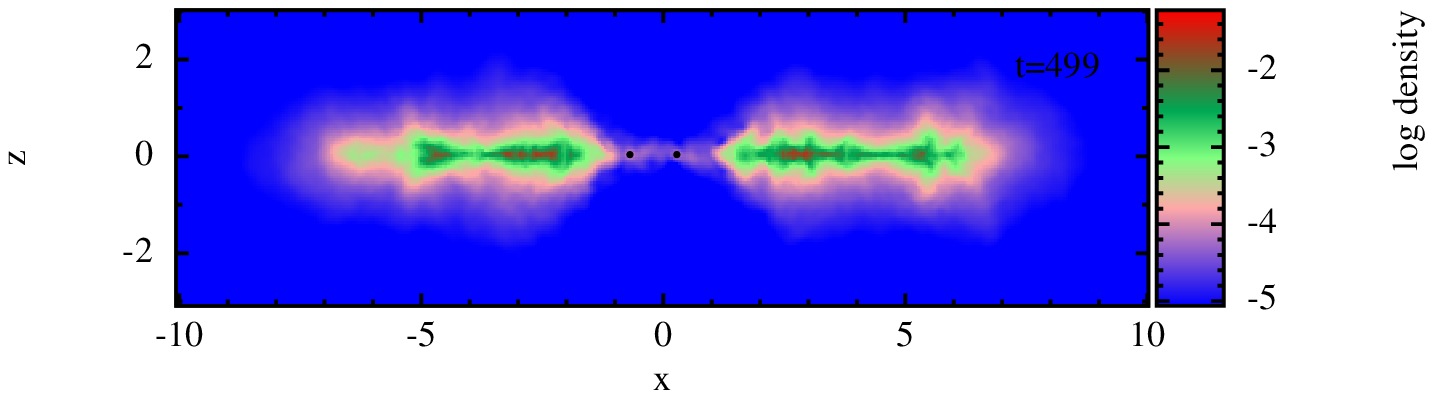}
 \includegraphics[width=\linewidth]{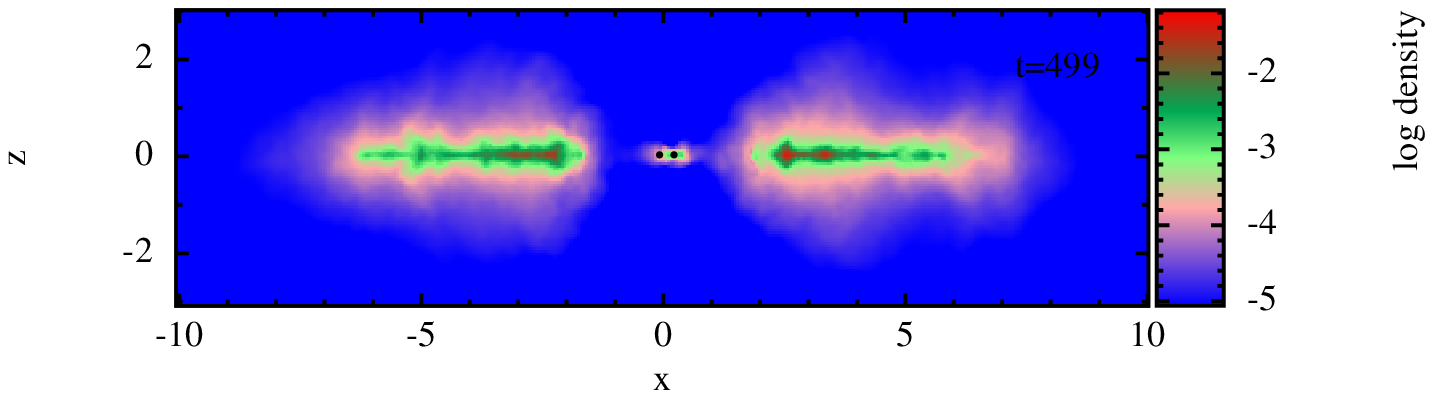}
\caption{Relaxed meridional density maps of the disc for the \textit{adia} (top) and \textit{iso} (bottom) runs; 
the black dots indicate the BHs, the axis are in units of the binary semi-major axis $a$. Note that this is not 
an azimuthal average but a vertical slice of the disc. \label{fig:initrho}}
\end{figure}

\begin{figure}
 \includegraphics[width=\linewidth]{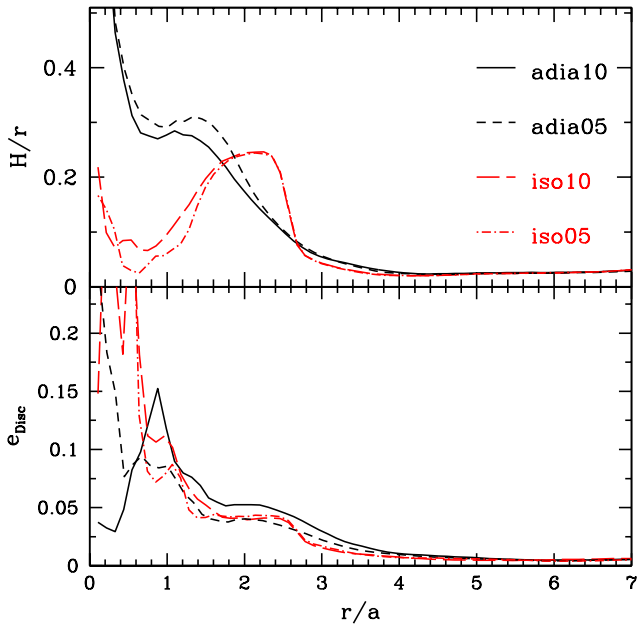}
\caption{Time and azimuth-averaged values of the disc scale-height $H/r$ and the disc eccentricity $e_{\rm disc}$ as a function of radius for the different simulations. \label{fig:H}}
\end{figure}

The model and numerical setup of this work is closely related to that of \cite{Roedig2011} and \cite{Jorge09}, 
hence we only outline the key aspects in the following, and refer the reader to these two papers for further details.

We simulate a self-gravitating gaseous disc of mass $M_d=0.2 \,M$
around two BHs of combined mass $M=M_1+M_2$, mass ratio
$q=M_2/M_1 = 1/3$, eccentricity $e$ and semi-major axis $a$, using the
SPH-code {\sc Gadget-2} \citep{Springel05} in a modified version that
includes sink-particles which model accretion on to the BHs
\citep{Springel05b, Cuadra2006a}.  Moreover, the orbit of the BHB is
followed very accurately by using a fixed small time-step and summing
up directly the gravitational force from every other particle in the
simulation \citep{Jorge09}.  The disc, which is co-rotating with the
BHs, radially extends to about $7a$, contains a circumbinary
cavity of radial size $\sim2a$ and is numerically resolved by 2 million\footnote{To obtain better resolution in the low density region inside the cavity,
we performed two shorter runs using 8 million particles, finding no
appreciable difference in any of the results discussed below.}
particles. The numerical size of each BH is denoted as $r_{\rm sink}$, the 
radius below which a particle is accreted, removed
from the simulation, and its momentum is added to the BH \citep{Bate95}. 
We do not scale $r_{\rm sink}$ by BH-mass, but use one fixed value
for both BHs. For all runs, the size of $r_{\rm sink}$ is smaller than the 
Roche lobe of both BHs, respectively. A particle is accreted if its separation from either
BH is smaller than $r_{\rm sink}$, no other conditions are imposed.
.
The gas in the disc is allowed to
cool on a time scale proportional to the local dynamical time of the
disc $t_{\rm dyn}=f_0^{-1}=2 \pi / \Omega_0$, where $\Omega_0 =
(GM_0/a_0^3)^{1/2}$ is the initial orbital frequency of the
binary\footnote{Subscripts $0$ refer to the initial values of any
  parameter.}.  To prevent it from fragmenting, we force the gas to
cool slowly, setting $\beta = t_{\rm cool}/t_{\rm dyn} = 10$
\citep{Gammie01,Rice05}.   The choice of $\beta$ is motivated by the 
requirement of maintaining the disc in a configuration of marginal
stability (Toomre parameter $Q$ in the range [1,2], see Section
\ref{sec:discstructure} and Fig. \ref{disc}). 
Assuming $\beta=10$, we thus effectively create an environment where
self gravity acts as a viscosity that can transport angular momentum 
outwards.

We use two different treatments for the
thermodynamics inside the cavity, denoted as \textit{adia} and
\textit{iso}, respectively. In the \textit{adia} runs, the gas within
the cavity is allowed to both cool via the $\beta$ prescription and to
heat up adiabatically, as it is the remainder of the disc
\citep{Jorge09}. In the \textit{iso} runs, we define a threshold
radius $r_{\rm cavity}= 1.75a$, below which the gas is treated
isothermally, meaning that the internal energy per unit mass is set to
be $u \approx 0.14 (GM/R)$ \citep{Roedig2011}. These two
numerical experiments were chosen to be highly idealised,
allowing us to study how torques behave in two opposite scenarios:
one in which hydrodynamics is important in the cavity -- the gas can 
cool efficiently and settles down in the binary plane forming 
mini-discs ({\it iso}); and one in which hydrodynamical forces have 
very little influence inside the cavity and the dynamics is dominated
by the gravity of the BHs (\textit{adia}).

\begin{table}
  \caption{Initial parameters and names of the runs in unitlengths of the code.
 \label{tab:parameters}}
\begin{center}
  \begin{tabular}{cccc}
    \hline \hline
Model & $r_{\rm sink} $ & Cavity EoS & Disc EoS \\
    \hline
$\mathtt{adia10}$   & $0.1$  & adiabatic+$\beta$& adiabatic +$\beta$\\
$\mathtt{iso10}$    & $0.1$  & isothermal&adiabatic +$\beta$\\
   \hline
$\mathtt{adia05}$   & $0.05$ & adiabatic+$\beta$&adiabatic +$\beta$\\
$\mathtt{iso05}$    & $0.05$ & isothermal& adiabatic +$\beta$\\
   \hline
    \hline
  \end{tabular}
\end{center}
\end{table}

As initial conditions we use a relaxed snapshot from \cite{Jorge09}
taken at their $t=500 \Omega_0^{-1}$ (see \cite{Roedig2011}, Section
2.1)\footnote{Notice that in the remainder of this article we refer to
  the time of this snapshot as $t=0$.} . The nomenclature of the runs
and the parameters used are listed in Tab.~\ref{tab:parameters}. Each
run is evolved for $\approx 90$ binary orbits, and we store the output
in single precision $\sim 6$ times per orbit.  We also closely track
accretion by storing the position and velocity of the two BHs and of
each accreted particle at the time the latter crosses the sink
radius. For both prescriptions (\textit{adia} and \textit{iso}) we
show the relaxed density configurations sliced along the meridional
plane in Fig.~\ref{fig:initrho} (see also
Figs.~\ref{disc_surfacedeniso}--\ref{disc_surfacedenadia} for a
face-on view)\footnote{Plots made using {\sc SPLASH}
  \citep{Price2007}.} and the measured, time-averaged values for both
the (semi) scale height of the 
disc $H/r$, and the 
eccentricity of the disc
$e_{\rm disc}$ in Fig.~\ref{fig:H}.
 $H$ is taken to be the height
above and below the disc-orbital plane in which $70\%$ of the mass
inside the annulus $(r,r+dr)$ is found.
Generally, we can
consider the physics to be resolved if the smoothing length 
$h$
\footnote{ We recall, that {\sc Gadget} defines $h=2\bar{h}$,
where $\bar{h}$ is the conventional definition of SPH smoothing 
length \citep{Springel05}. Therefore, all the numbers we give should
be divided by two to get the resolution in the conventional SPH language.
} 
is smaller than the characteristic lengthscale: radially the criterion
$h/r \ll 1$ is achieved well: $h/r \in (0.005,0.2)$; whereas
vertically 
the criterion $h< (c_s/\Omega)$ is well
satisfied throughout the simulation domain, except very close 
to the primary hole: $h/(c_s/\Omega)\in (0.1,0.5)_{|r\ge0.5}$.
Finally, our resolution ensures that $h\lsim r_{\rm sink}$ close
to the BHs, a condition which is necessary to properly resolve accretion
onto a sink particle.

Unless otherwise stated, we will present all the relevant quantities
in the natural units of the simulation by setting $G=M_0=a_0=1$. It
follows that also the initial circular velocity of the binary is
$V_{c,0}=1$ and its initial period $P_0=2\pi$. We will then discuss the
astrophysical implications of our findings by scaling them to fiducial
astrophysical BHB systems.

\section{Binary evolution: causes}
\label{sec:evolution}

\begin{figure}
\centering
\includegraphics[width=\linewidth]{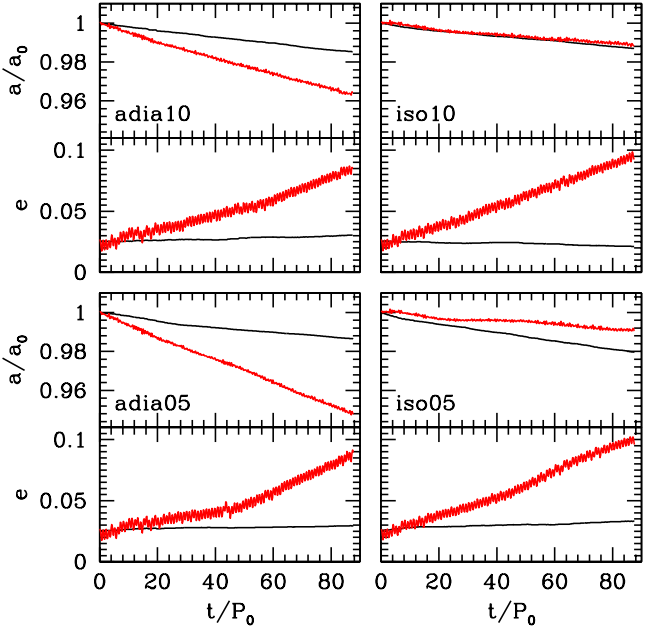}
\caption{Semimajor axis (top panel in each plot) and eccentricity (bottom panel in
each plot) evolution for all the four runs. In each panel, the \textcolor{red}{red} 
line represents the full evolution directly taken from the simulation whereas the \textcolor{black}{black} line
is the equivalent evolution when considering the energy and angular momentum exchanges
due to accreted particles only (i.e., basically due the evolution of the mass and mass ratio of the binary).}
\label{delta_ae}
\end{figure}

For each of the four runs, we measure the two primary orbital
elements: eccentricity $e$ and semi-major axis $a$. Their evolution
is shown by the \textcolor{red}{red} lines in Fig.~\ref{delta_ae}.  In
all four runs, we find $\dot{a}(t)< 0$ and $\dot{e}(t)> 0$. While the
eccentricity evolution is largely independent on the sink radius value
and the adopted EoS within the cavity, the orbital shrinking is much
faster in the {\it adia} runs, in which the binary shrinks by $4-5\%$
over 90 orbits. Conversely, in the {\it iso} runs, the two BHs get
only about $1\%$ closer. Linear extrapolation of such 
 instantaneous shrinking rates at face value  would imply 
binary coalescence on a time-scale of $\sim 2000 P_0$ and
$\sim 10^4 P_0$ for the {\it adia} and the {\it iso} runs,
respectively. Whereas, assuming a constant eccentricity growth rate,
the limiting value predicted by \cite{Roedig2011} would be reached in
$\lsim 1000 P_0$. 
As a note of caution, for a very similar numerical setup, \citet{Jorge09} 
showed that the secular evolution of $a$ and $e$ departs from being 
linear for longer timescales.  This has to be expected given the 
limited energy and angular momentum reservoir provided by an isolated 
circumbinary disc, which cannot transfer further out the angular 
momentum extracted by the binary.
However, in a more realistic situation (e.g., a binary interacting with a steady inflow of gas), such slow- down may well not happen.
How the long-term secular evolution is influenced by
external replenishing of the disc is beyond the scope of this paper.

\subsection{Gravitational torque and accretion contribution to the angular momentum budget}
Being interested in the physical mechanisms driving the binary
evolution, we firstly identify two distinct processes:
\begin{enumerate}
\item the gravitational torques exerted by the gaseous particles onto
  each individual BH, ${\bf T}_{\rm G}$ (gravity torque)
\item the accretion of instreaming particles crossing either BH sink radius,
  $(d{\bf L}/dt)_{\rm acc}$ (accretion torque){\footnote{This is similar
to the angular momentum transfer due to accretion at the inner excision
boundary defined by S11.}}.
\end{enumerate}
Conservation of the total angular momentum in the simulations implies
\begin{equation}
\frac{d{\bf L}}{dt}={\bf T}_{\rm G} + \frac{d{\bf L}}{dt}_{\rm acc}, 
\label{dl}
\end{equation}
where ${\bf L}$ is the BHB orbital angular momentum vector 
\footnote{Throughout the paper we will denote all vectorial quantities in
boldface.}.  All vectors are computed with respect to a Cartesian
reference frame centred in the BHB centre of mass (CoM){\footnote{The
    binary CoM slightly wiggles around the total binary--disc CoM. As a
    cross check, we also computed all the relevant quantities with
    respect to the latter, finding no significant difference in the
    results.}}; the binary initially lies in the $x$--$y$ plane with
angular momentum oriented along the positive $z$ axis. In our SPH
simulations, ${\bf T}_{\rm G}$ can be computed at each snapshot by direct
summation of individual particle torques onto each BH yielding
\begin{equation}
{\bf T}_{\rm G}=\sum_{j=1}^N\sum_{k=1}^2 {\bf r}_k\times\frac{GM_km_j({\bf r}_j-{\bf r}_k)}{|{\bf r}_j-{\bf r}_k|^3},
\label{torque}
\end{equation}
where $j$ runs over all $N$ gas particles, $k$ identifies the two BHs, ${\bf r}$ 
are position vectors, and $m$ and $M$ denote particle and BH masses respectively. 
On the other hand, $(d{\bf L}/dt)_{\rm acc}$ can be computed by assuming instantaneous 
linear momentum conservation of each accreted particle yielding 
$\Delta {\bf L}={\bf r}_{k,j}\times m_j{\bf v}_j$. Here ${\bf r}_{k,j}$  is
the position vector of the accreting BH at the moment of swallowing the particle $j$, 
and ${\bf v}_j$ is the particle velocity vector.
Over the interval between two subsequent snapshots,
we evaluate the binary angular momentum change, $\Delta{\bf L}$,
by splitting Eq.~(\ref{dl}) into two parts: 
\begin{equation}
\Delta{\bf L}={\bf T}_{\rm G} \Delta{t}+\sum_{\Delta{t}} {\bf r}_{k,j}\times m_j{\bf v}_j,
\label{dlnum}
\end{equation}
where $\Delta{t}$ is the interval between the two snapshots, and the sum runs 
over all the accretion events $j$ occurring during this time lapse at their respective
positions. Note, that we
assume an accretion event to be instantaneous, i.e. the accreting BH $k$ does not move during
the accretion.

\begin{figure*}
\begin{tabular}{cc}
\includegraphics[scale=0.4,clip=true,angle=0]{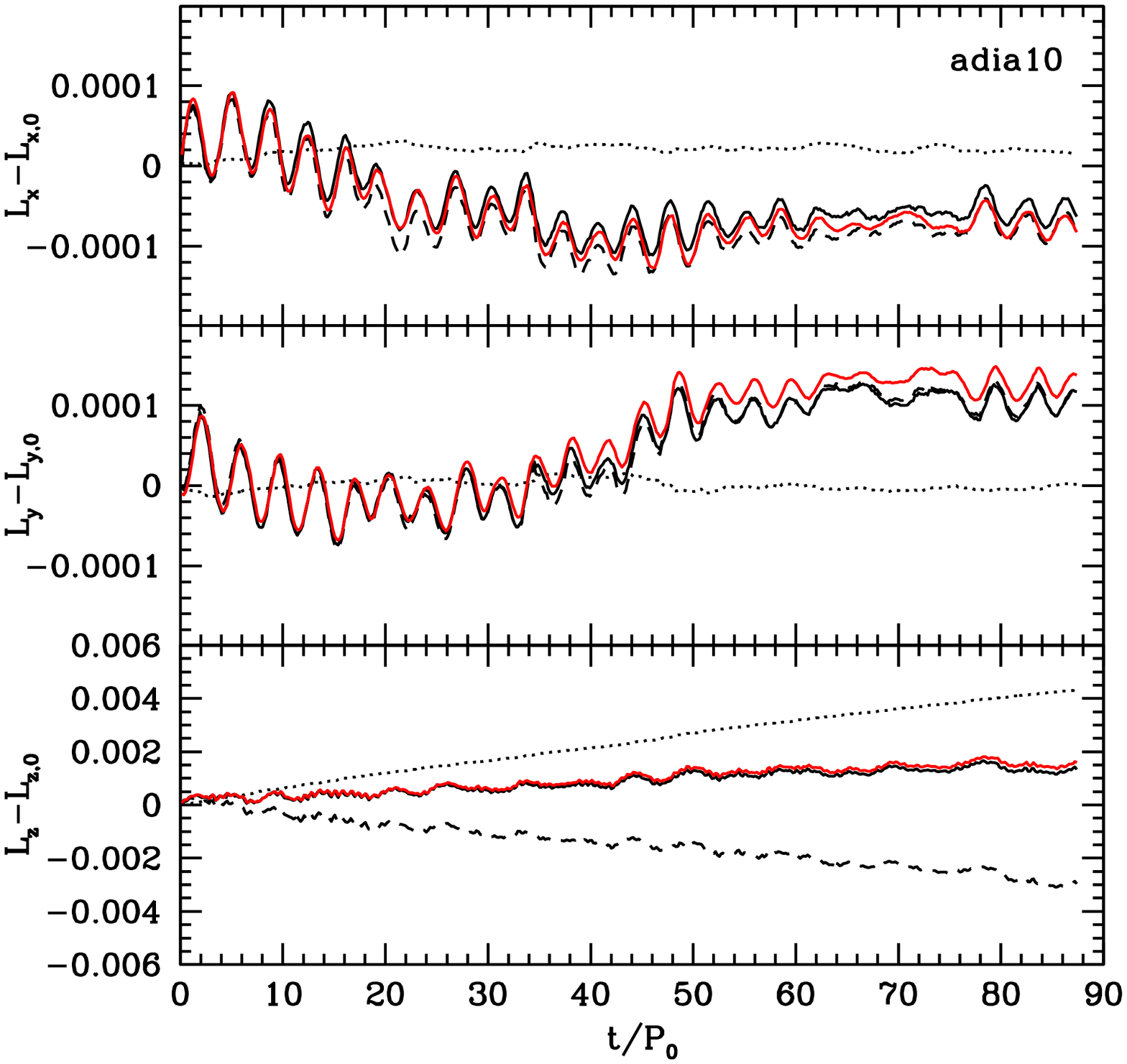}&
\includegraphics[scale=0.4,clip=true,angle=0]{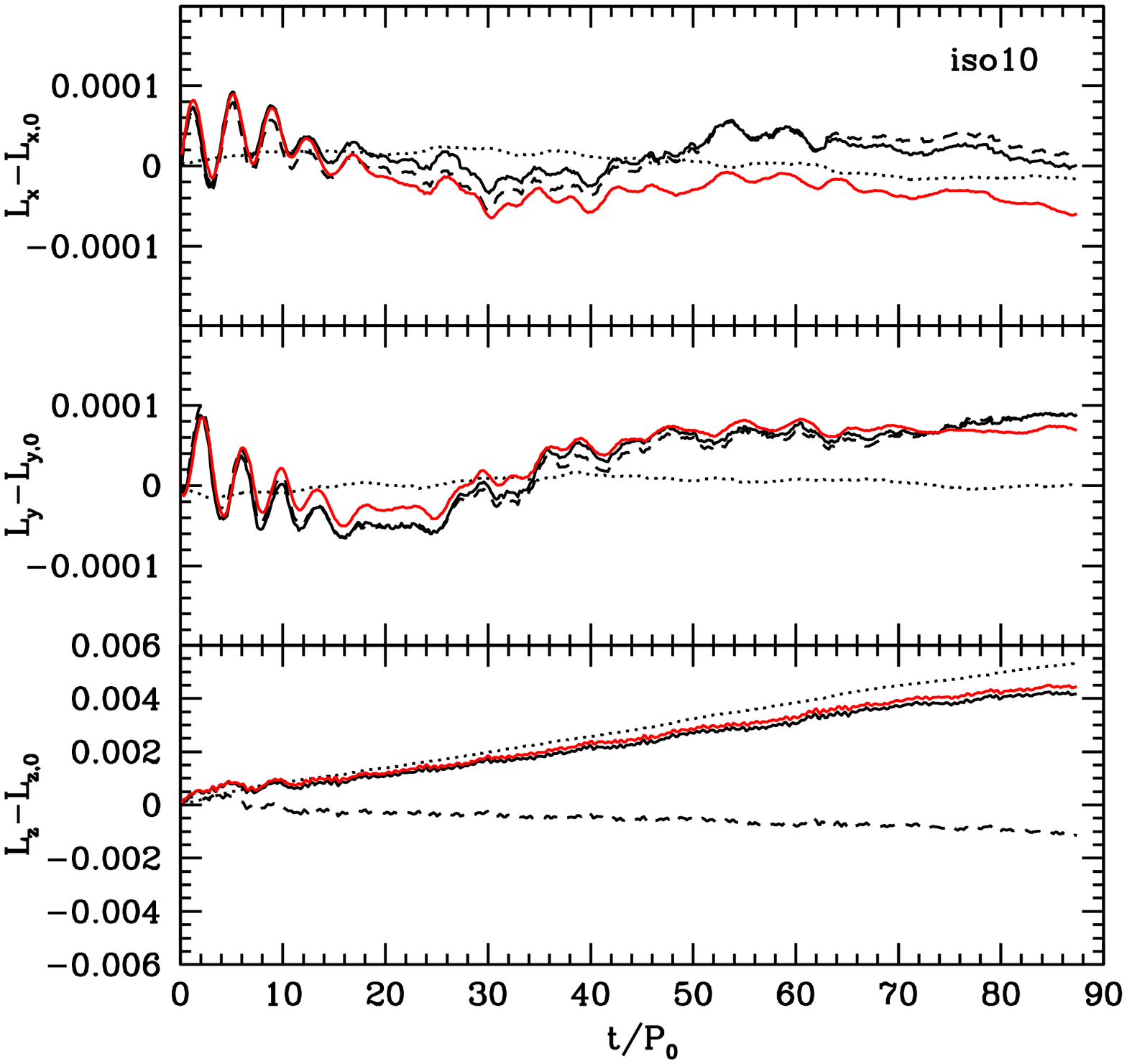}\\
\includegraphics[scale=0.4,clip=true,angle=0]{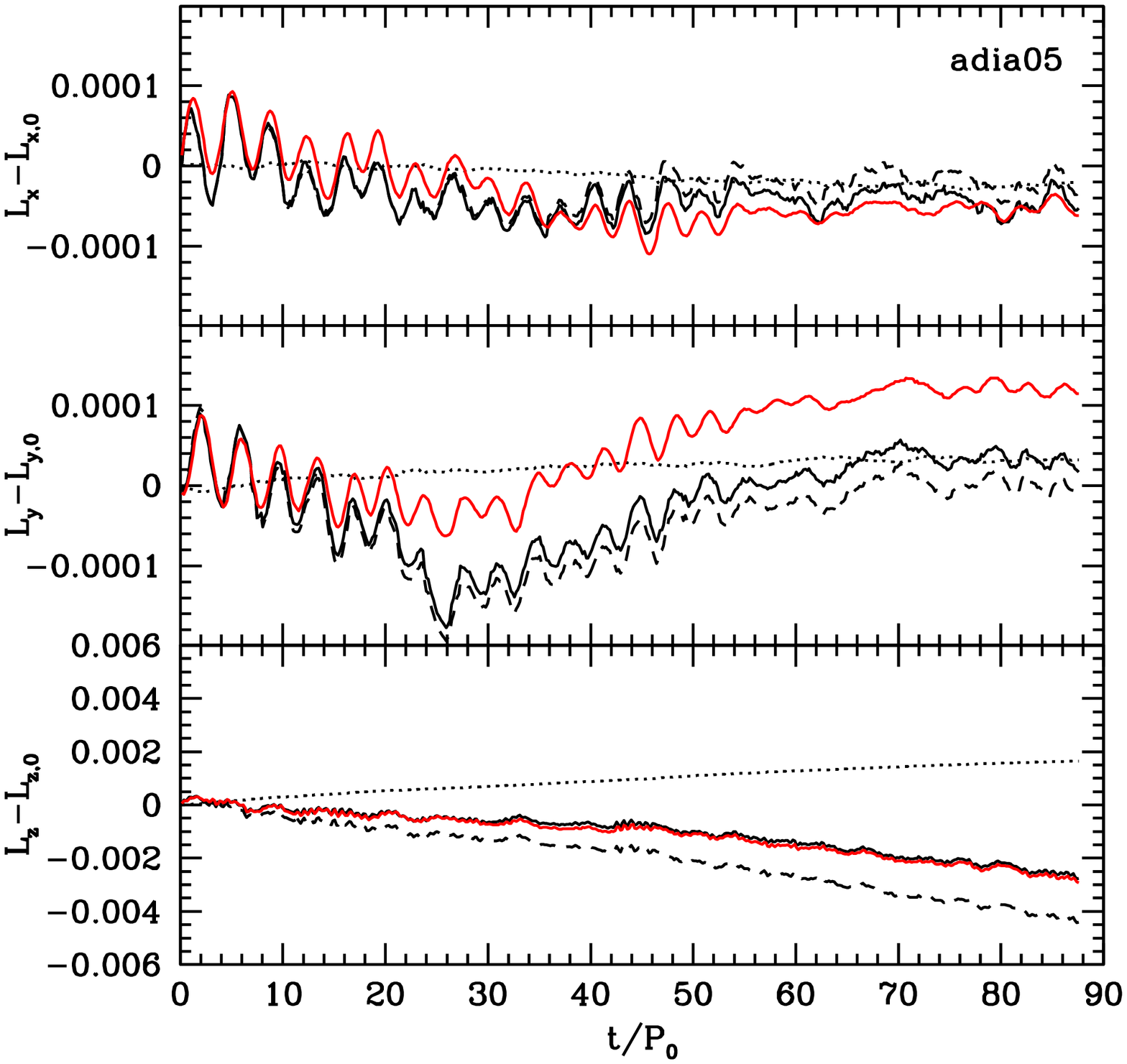}&
\includegraphics[scale=0.4,clip=true,angle=0]{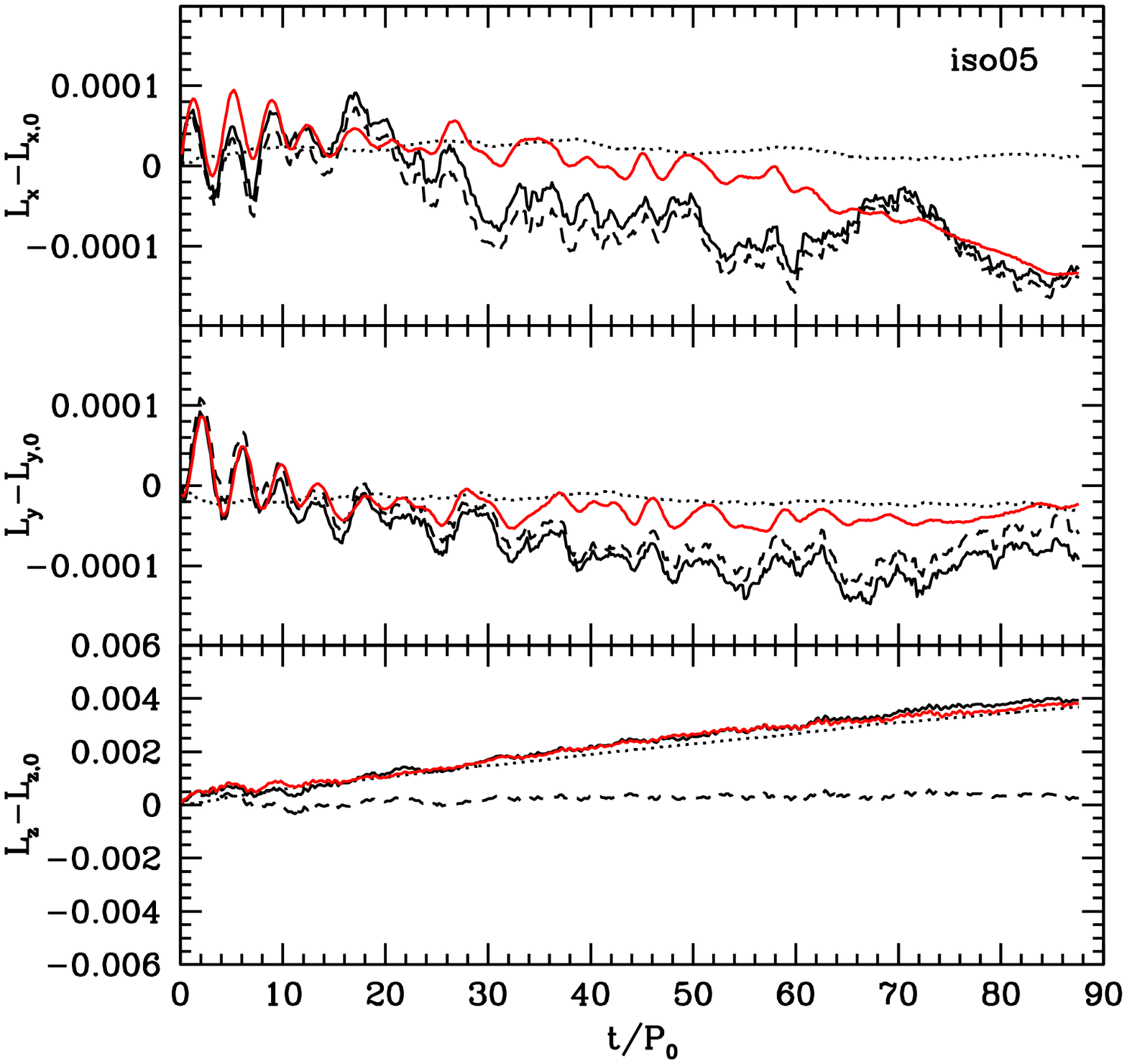}\\
\end{tabular}
\caption{Consistency check of the evolution of the binary angular momentum components for
all the simulations. In each plot, the $L_x$, $L_y$ and $L_z$ component evolution is shown
from the top to the bottom. In each panel, the dashed line is the $\Delta L$ exchange
balancing the gravity torques at each time step, the dotted line is the $\Delta L$ exchange
computed from the accreted particles, and the solid black line is the sum of those two.
For comparison, the \textcolor{red}{red} line is the angular momentum evolution taken directly from the 
raw simulation data. All $\Delta L$ are in units of $[M_0a_0V_{c,0}]$.}
\label{lz_evol}
\end{figure*}
 
We use Eq.~(\ref{dlnum}) to assess the relative importance of gravitational torques
and accretion in the evolution of the system. In Fig.~\ref{lz_evol}, we plot
the evolution of the $x,y,z$ components of the angular momentum 
${\bf L}$ as directly evaluated from the positions and velocities of the two 
BHs stored in the simulation outputs (\textcolor{red}{red}), together with the 
relative changes due to accretion (black dotted) and gravity torques (black dashed) 
according to Eq.~(\ref{dlnum}).
The solid black line is the sum of the latter
two which should overlap with the red line if our decomposition is
sufficiently accurate and these are the only two physical processes
determining the binary evolution. For comparison, in simulation units, the initial
binary angular momentum is $\approx 0.18$. The $L_x$ and $L_y$ components are
almost constant, showing fluctuations at the $10^{-4}$ level. The
mismatch between the \textcolor{red}{red} and the
\textcolor{black}{solid black} lines here is because simulation
outputs are in single precision, i.e. accurate to the fourth digit,
which is the fluctuation level of the computed quantity. On the
contrary, the $L_z$ component changes up to a $10^{-2}$ absolute
level (i.e., about $5\%$ of the initial value), and in this case we 
see very good agreement between the two lines.  
Note that in three of the runs, the overall angular momentum
grows over time, even though $a$ shrinks and $e$ increases. 
This counter intuitive result is due to the dominance of
accretion in the angular momentum budget, and will be further
discussed in \S~\ref{sec:dissection} below (see also S11).

As a general rule, the accretion contribution to the binary ${\bf L}$
evolution 
is at least comparable to the effect of the gravitational
torques. It is therefore also interesting to see the accretion
contribution to the evolution of the binary orbital elements. Having
stored all the accretion events, this can be done separately by simply
evolving the binary from $a_0$ and $e_0$ only by adding one accreted
particle after the other, imposing linear momentum conservation. This
"\textit{would-be}" evolution of $e(t)$ and $a(t)$, accounting only
for the accretion effect, is shown in \textcolor{black}{black} in all
panels of Fig.~\ref{delta_ae}.  It is clear that $e(t)$ is mainly
driven by gravitational torques, with accretion playing a negligible
role.  This is in line with our previous findings \citep{Roedig2011},
where the evolution of the eccentricity was attributed mainly to the
gravitational interaction between the binary and the overdensities
excited by its quadrupolar potential in the inner rim of the
circumbinary disc. Conversely, the $a(t)$ plots highlight the
importance of accretion for the semimajor axis evolution. In the
\textit{iso05} case, the binary shrinking due to accretion is actually
larger than the total one, meaning that the disc torques would force
the binary to expand; an effect similar to that described by
\cite{LinPapa11} in the context of planetary migration in
self-gravitating discs. Note that in the {\it adia} cases $\dot{a}$ is
dominated by the effect of the disc.  This explains the match between
such simulations and the \cite{Ivanov99} analytical model based on
angular momentum transport through the disc \citep{Jorge09}.

\subsection{Dissection of the binary evolution into its components}
\label{sec:dissection}
\begin{figure}
\includegraphics[width=\linewidth]{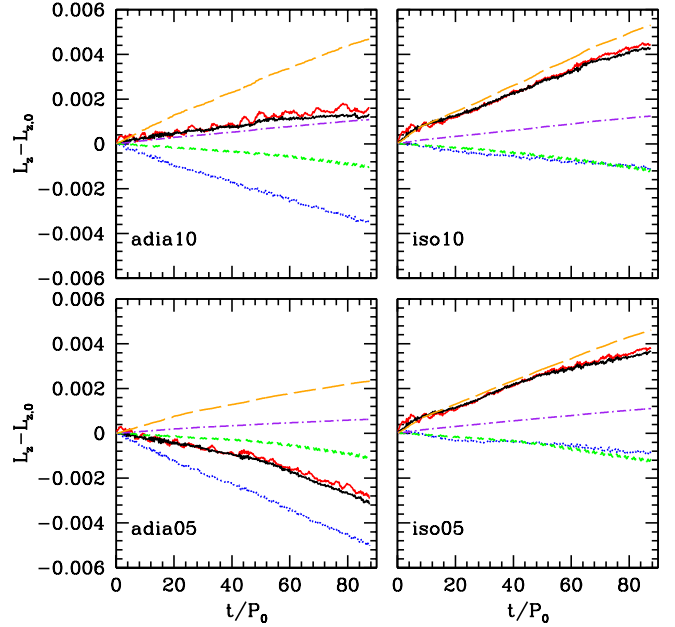}
\caption{Decomposition of the $L_z$ evolution for the four runs. The $\Delta{L_z}$ contributions stem from 
\textcolor{blue}{$a$ (dotted)}, \textcolor{green}{$e$ (dashed)}, \textcolor{Plum}{$M$ (dot dashed)} 
and \textcolor{BurntOrange}{$\mu$ (long dashed)} variations separately. The sum of the four contributions 
is given by the solid black line, whereas the solid \textcolor{red}{red} line is, as in Fig. \ref{lz_evol}, 
the angular momentum evolution taken directly from the raw simulation data. All $\Delta L$ are in units 
of $[M_0a_0V_{c,0}]$.}
\label{decomp}
\end{figure}

So far, we have separated the relative contribution of gravitational
torques and accretion to the binary angular momentum budget. We now
investigate how the angular momentum change is distributed among the
relevant binary quantities. As clearly shown in Fig.~\ref{lz_evol},
$L_x$ and $L_y$ show only small fluctuations (and therefore remain 
small compared to $L_z$, since the binary initially orbits in the $x$--$y$ 
plane). From here on, we will therefore concentrate on the dominant $L_z$
component. The binary angular momentum is
\begin{equation}
L_z=\mu\sqrt{GMa(1-e^2)},
\label{eqlz}   
\end{equation}
where $\mu=M_1M_2/M$ is the binary reduced mass. Eq.~(\ref{eqlz}) can be differentiated to separate 
the contribution of each single relevant quantity to the angular momentum change:
\begin{equation}
\frac{\dot{L}_z}{L_z}=\frac{\dot{a}}{2a}+\frac{\dot{M}}{2M}+\frac{\dot{\mu}}{\mu}-\frac{e}{1-e^2}\dot{e}.
\label{eqlzdec}   
\end{equation}
By directly measuring \textcolor{blue}{$a$}, \textcolor{green}{$e$}, \textcolor{Plum}{$M$}, \textcolor{BurntOrange}{$\mu$} from the simulation, we can evaluate each single term and decompose the 
$L_z$ evolution accordingly. This is shown in Fig.~\ref{decomp} for all the four runs. Note that the sum of the 
four contributions (solid black line in each panel) closely matches the measured $L_z$ evolution directly measured 
from the simulation (\textcolor{red}{red}), validating our first-order expansion.  

It is worth noting that 
the increase of $L_z$ is perfectly compatible with semi-major axis shrinkage and eccentricity growth. 
In fact, Eq.~($\ref{eqlzdec}$) can be inverted to obtain 
\begin{equation}
\frac{\dot{a}}{a}=\frac{2T_z}{L_z}-\frac{\dot{M}}{M}-\frac{2\dot{\mu}}{\mu}+\frac{2e}{1-e^2}\dot{e},
\label{adotdec}   
\end{equation}
where we used the fact that $\dot{L}_z\equiv T_z$. Even if the total torque
is positive, and the eccentricity grows, the binary can shrink because
of the negative contribution of the $M$ and the $\mu$ terms. In
particular, as we will see below, the higher accretion onto the
secondary hole results in a large increase of $\mu$ (long-dashed line
in Fig.~\ref{decomp}) which is the main driver of the binary angular
momentum growth.  Eqs.~(\ref{eqlzdec}) and~(\ref{adotdec}) highlight
the importance of accretion in determining the binary temporal
evolution.

\section{Gravitational torques}
\label{sec:torx}
As pointed out in the previous Section, understanding secular dynamics
of BHBs in gaseous environments is closely tied to studying the
interplay between long distance gravitational forces, hydrodynamics
and accretion. In this Section, we focus on the torques coming from
the self-gravitating gaseous disc, investigating the influence of
regions at different distances from the BHB.

\subsection{Time averaged torque profiles}
Gravitational torques exerted by the disc onto the BHB can be directly
calculated at each snapshot from Eq.~(\ref{torque}).  It is of great
interest to understand where such torques originate, and given the
cylindrical nature of the problem, it is natural to investigate their
azimuthally-averaged radial distribution.  We take $r$ to be the
projected radial coordinate (in the $x$--$y$ plane) from the binary
CoM, and we define $dT/dr$ to be the differential torque,
integrated over the azimuthal coordinate and the disc height. The
total average torque exerted by gas in the projected distance interval
$[a,b]$ from the binary CoM is
\begin{equation}
\langle T_{[a,b]}\rangle=\left\langle\int_a^b\frac{dT}{dr}dr\right\rangle,
\label{eqavtorque}   
\end{equation}
where $\langle\cdot\rangle$ denotes temporal averaging over the entire simulation.
 
\begin{figure}
\includegraphics[width=\linewidth]{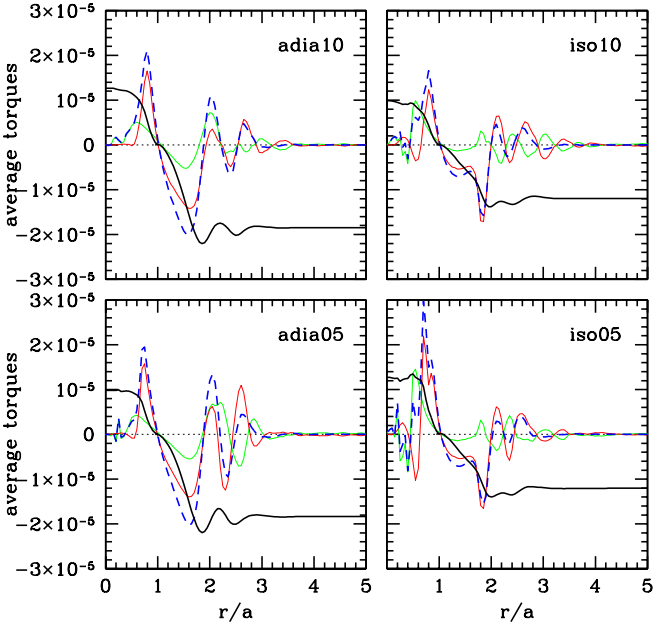}
\caption{Differential torques $dT/dr$ in units of $[G M^2_0a^{-2}_0]$ averaged over the entire simulations.  
In each panel we show the differential torque on the primary (\textcolor{green}{green}), on the secondary 
(\textcolor{red}{red}), the sum of the two (\textcolor{blue}{blue}), and the total integrated torque 
(\textcolor{black}{black}) according to Eq. \ref{eqavtorque}. This latter is integrated starting 
from $a$ inwards and outwards. Notably, the inward torque is positive, whereas the outward torque is 
negative. }
\label{torqueave}
\end{figure}

In Fig.~\ref{torqueave}, we show the time averaged differential torque acting on the binary, 
$\langle dT/dr\rangle$ 
(\textcolor{blue}{blue}), together with its integral according to Eq.~(\ref{eqavtorque}) (\textcolor{black}{black}). 
We also decompose the former in the two 
components acting on the primary (\textcolor{green}{green}) and the secondary (\textcolor{red}{red}) BH. All torques 
are null at the binary corotation radius 
 we therefore show the total torque by integrating from this point inwards 
($\langle T_{[1,0]}\rangle$, binary region) and outwards ($\langle T_{[1,\infty]}\rangle$, disc region). 

In all simulations, the local average torque shows an oscillatory
behaviour with a sharp maximum at the location of the secondary BH,
${r}\sim0.75$, and a deep minimum in the cavity region at $1<{r}<2$. In
the body of the disc (${r}>2$) positive and negative peaks alternate,
but they are shallow and almost cancel out, giving a negligible
contribution to the total torque (as witnessed by the fact that the
integrated torque is basically constant for ${r}>2$).  Note that the
torque on the secondary BH is always larger than the torque on the
primary, due to its proximity to the outer disc resulting in a
stronger interaction. The general behaviour is qualitatively the same
in the {\it adia} and {\it iso} runs. In this latter case, however, we
found a much sharper negative peak at ${r}\approx 1.7$ followed by a
smaller secondary negative bump at ${r}\approx1.2$. This appears to be
an artifact of the sudden change in EoS of the gas inside the cavity, at
${r}=1.75$. The net result is a smaller negative $\langle
T_{[1,\infty]}\rangle$ which has important consequences on the binary
evolution. We see in fact that in the {\it iso} runs $\langle
T_{[1,0]} \rangle$ and $\langle T_{[1,\infty]}\rangle$ almost cancel
out, meaning that, overall, gravitational torques do not change the
binary angular momentum.  Conversely, in the {\it adia} runs $\langle
T_{[1,0]}\rangle$ is much smaller (in absolute value) than $\langle
T_{[1,\infty]}\rangle$, implying an efficient angular momentum
transfer from the binary to the gas \citep[MM08;][]{Jorge09}.

\subsection{Spectral analysis}
We now consider the torque evolution in time. We separately discuss torques
coming from different disc regions by showing both the time series and their associated power spectra. 
We cut the spatial domain into the following radial annuli: 

\begin{tabular}{lll}
(i)     &   $0\,\,\,\,{a}<r<10\,\,{a}$&: the entire domain \\
(ii)    &   $0\,\,\,\,{a}<r<1\,\,\,\,\,{a}$&:   the 'binary region' \\
(iii)   &   $1\,\,\,\,{a}<r<1.8{a}$&: the 'cavity region' \\
(iv)    &   $1.8{a}<r<2.5{a}$&: the 'rim region' \\
(v)     &   $2.5{a}<r<10\,\,{a}$&: the 'disc region' \\
\end{tabular}

\begin{figure*}
\begin{tabular}{cc}
\includegraphics[scale=0.4,clip=true,angle=0]{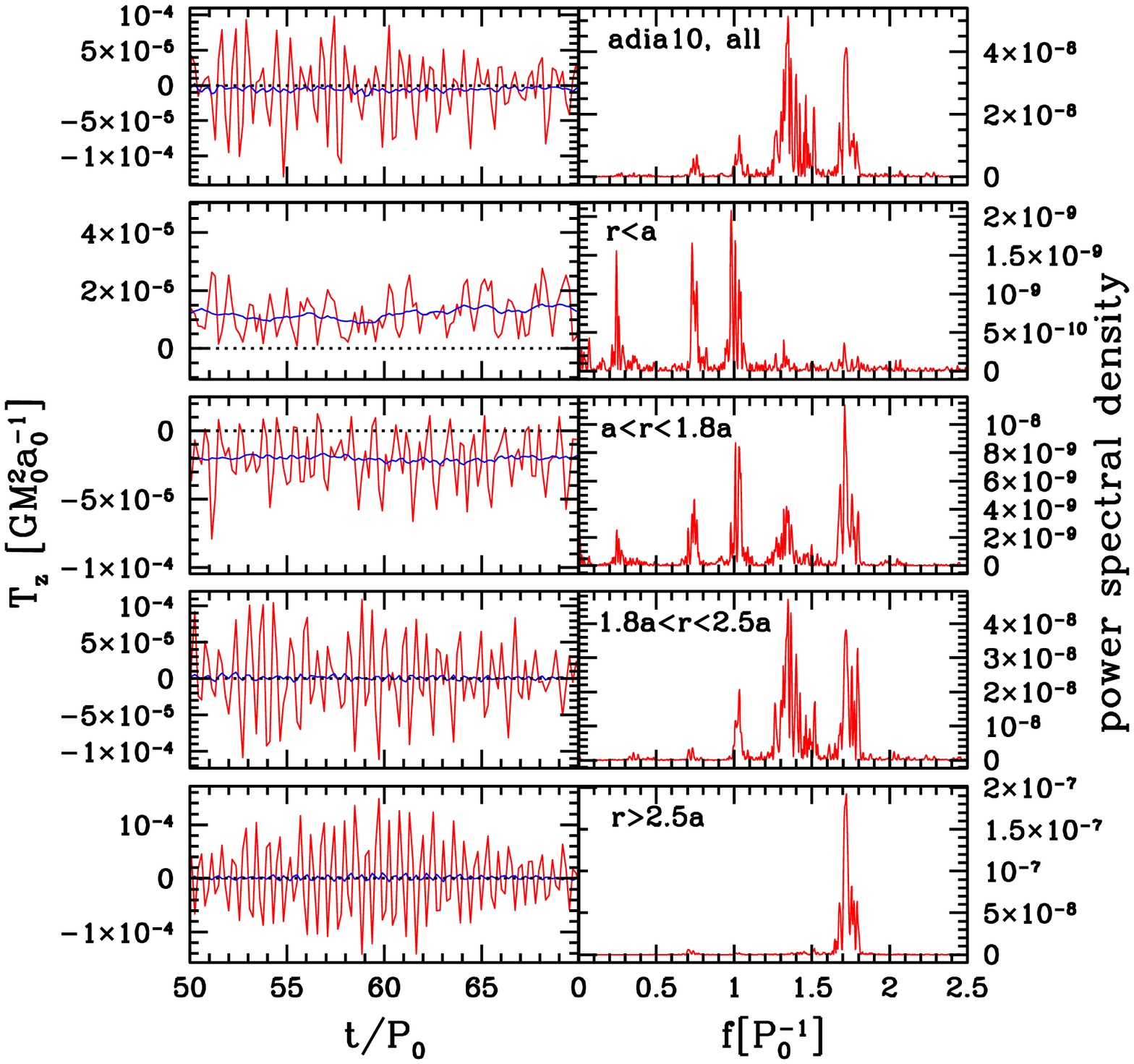}&
\includegraphics[scale=0.4,clip=true,angle=0]{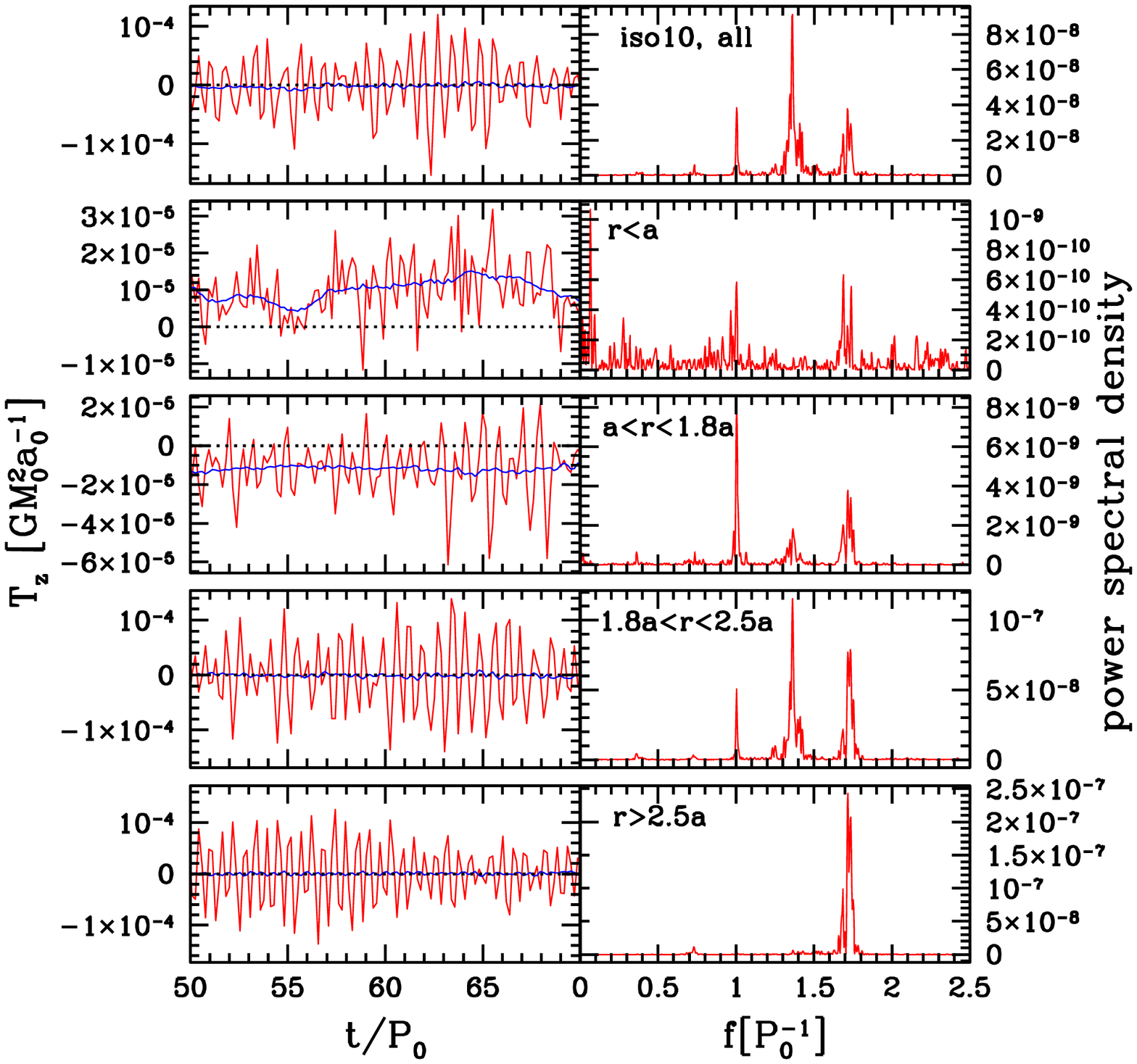}\\
\includegraphics[scale=0.4,clip=true,angle=0]{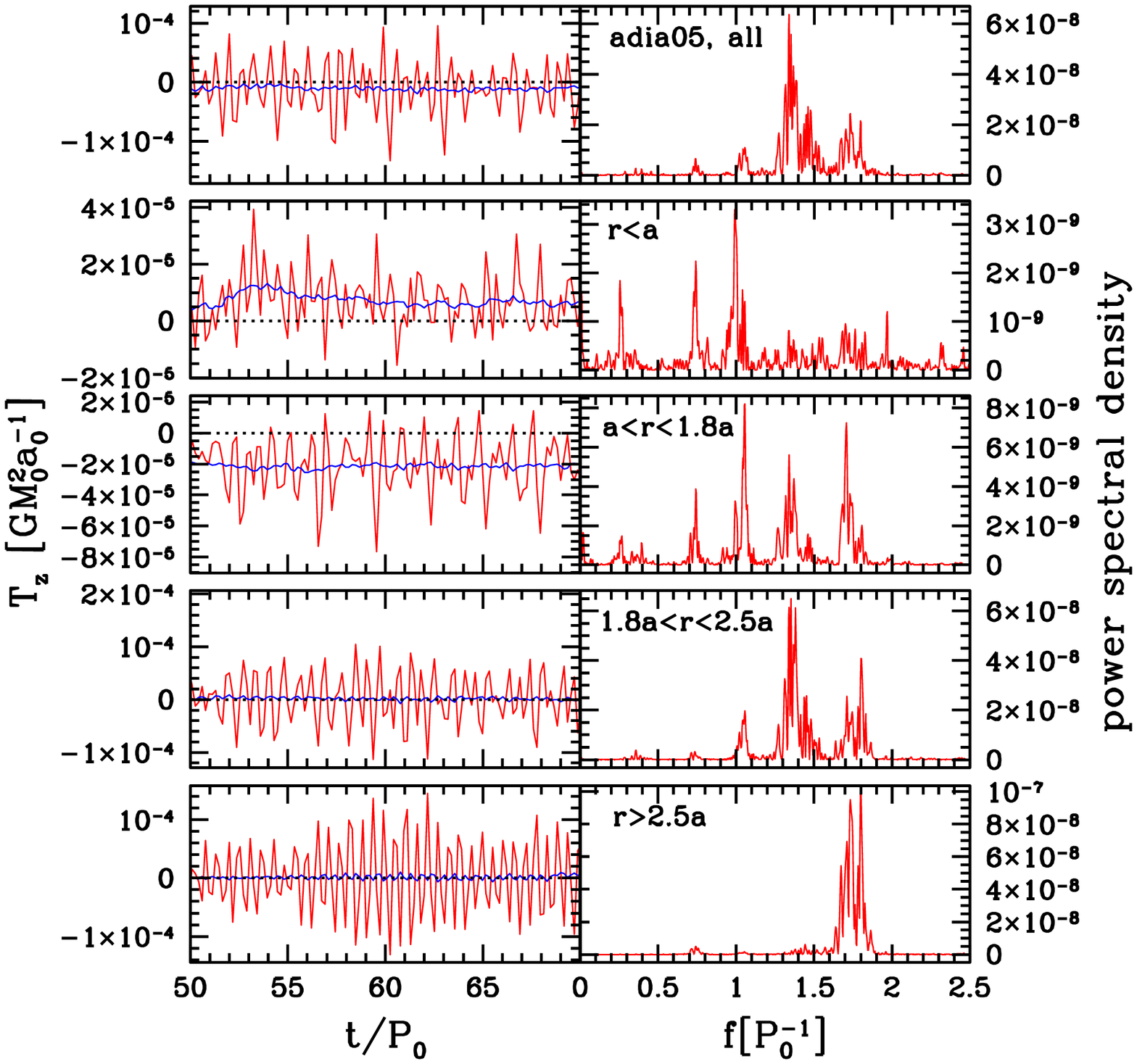}&
\includegraphics[scale=0.4,clip=true,angle=0]{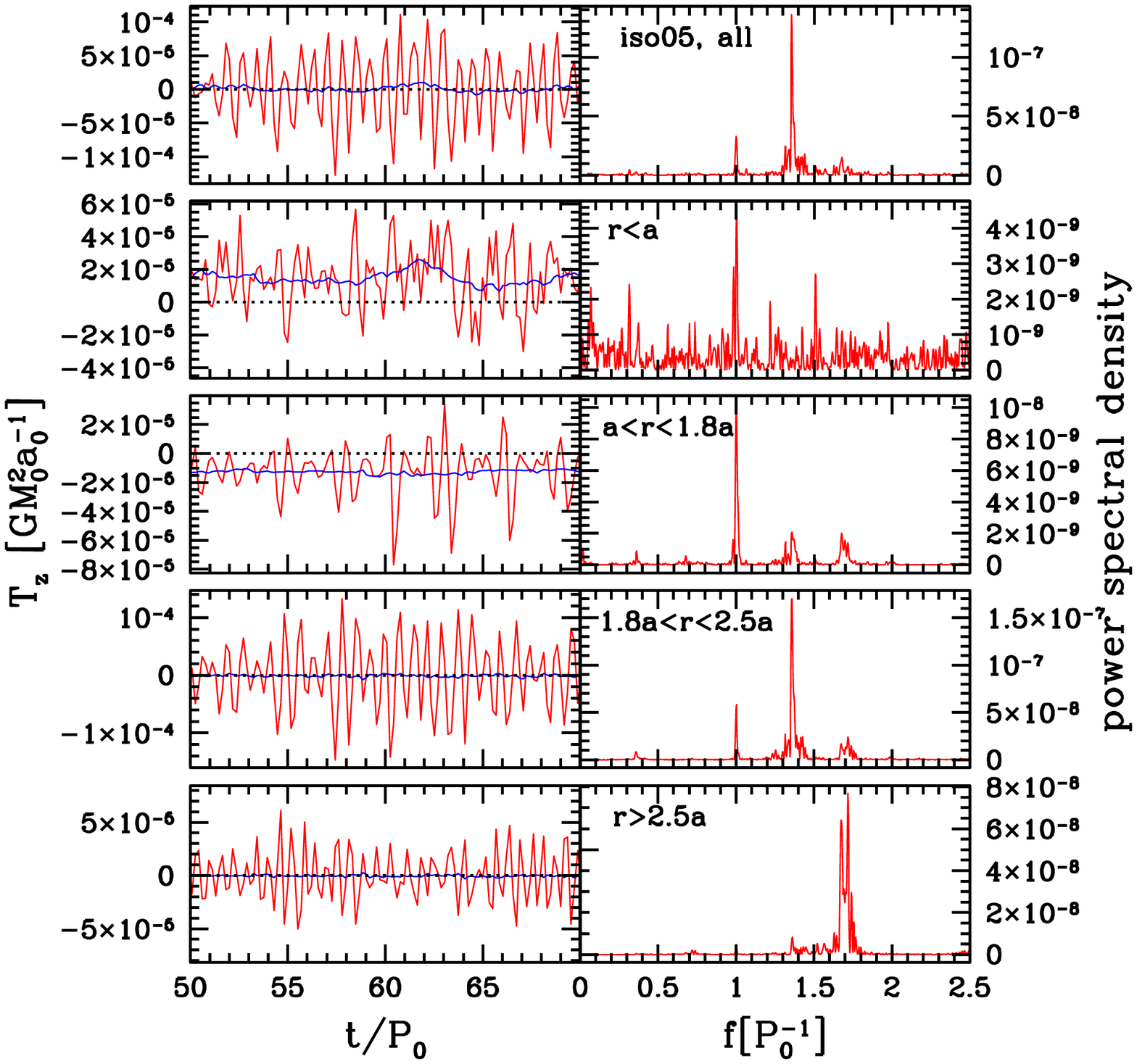}\\
\end{tabular}
\caption{Each plot depicts the torques exerted by the disc on the BHB in the time domain
(left panels) and in the frequency domain (right panels). In each plot, from the top to the 
bottom, pairs of panels refer to the five domains discussed in the text: total torque (i),  
torque exerted by the material located at $r<a$ (ii), $a<r<1.8a$ (iii), $1.8a<r<2.5a$ (iv) and $r>2.5a$ (v). 
In each left panel, the red line is the raw oscillating torque, and the blue one is the torque smoothed 
over three periods to show the average behaviour. In the associated right panel, we plot the
power spectrum as a function of frequency in units of the initial binary orbital frequency. 
All torques strongly oscillate, showing several characteristic frequencies, however, note the 
different scales of the panels.
}
\label{torquetotnew}
\end{figure*}

The associated time series and power spectra are shown respectively 
in the left and right panels of Fig.~\ref{torquetotnew}.
 
In all the simulations, the overall torque (i), shows a clear periodic
oscillation, much larger in amplitude than its average value. The
power spectrum unveils several distinctive peaks, with relative
amplitudes that can vary significantly for different simulations. In
particular, peaks in the {\it iso} simulations are much sharper and
better defined. This is because in the {\it adia} simulations the BHB
shrinks significantly, resulting in a broadening of the characteristic
frequencies. Moreover, as we shall see in the next section, the disc
sub-structures are much better defined in the {\it iso} runs, giving
rise to neater features. 

In the binary region (ii) the torque is
mostly coherent and positive. Because the torque strength in (ii) is
regulated by the mass inflow, it shows periodicities that are related
to the accretion flow: a disc component around $f\approx0.25 P_0^{-1}$
(corresponding to the disc peak density at $r\approx 2.5a$), the
forcing frequency of the binary at $f=P_0^{-1}$, and the beat between
these two at $f\approx0.75 P_0^{-1}$ (see Section~\ref{sec:spin} and
\cite{Roedig2011} \S~5.1).  In the cavity region (iii) the torque is
negative on average but strongly oscillating. Several periodicities
are detectable, the most striking being a peak at $f\approx P_0^{-1}$
which appears again to be directly related to the binary
period. 
In the rim region (iv) the torque is highly oscillating, and the strongest feature is a sharp peak 
at $f\approx 1.3 P_0^{-1}$, whereas in the disc region (v) the only significant spectral component 
is at $f\approx 1.7 P_0^{-1}$.
As a general trend, moving from the inner region to the disc body, torques become incoherent 
(i.e. they average to zero) and strongly oscillating (compare the power spectra scales 
in the different panels of each plot).

\subsection{Interpretation: torque origin and location in the disc}
In this Section, we provide a global interpretation of the features observed both in the radial 
distribution of the time averaged torques and in their temporal evolution.  
The arguments discussed below are supported by Figs.~\ref{disc_surfacedeniso}--\ref{disc_surfacedenadia}--\ref{disc} 
and by Tab.~\ref{tab:massflus}.

\subsubsection{Origin of the positive and negative peaks}
\begin{figure}
\centering
\includegraphics[scale=0.3,clip=true,angle=0]{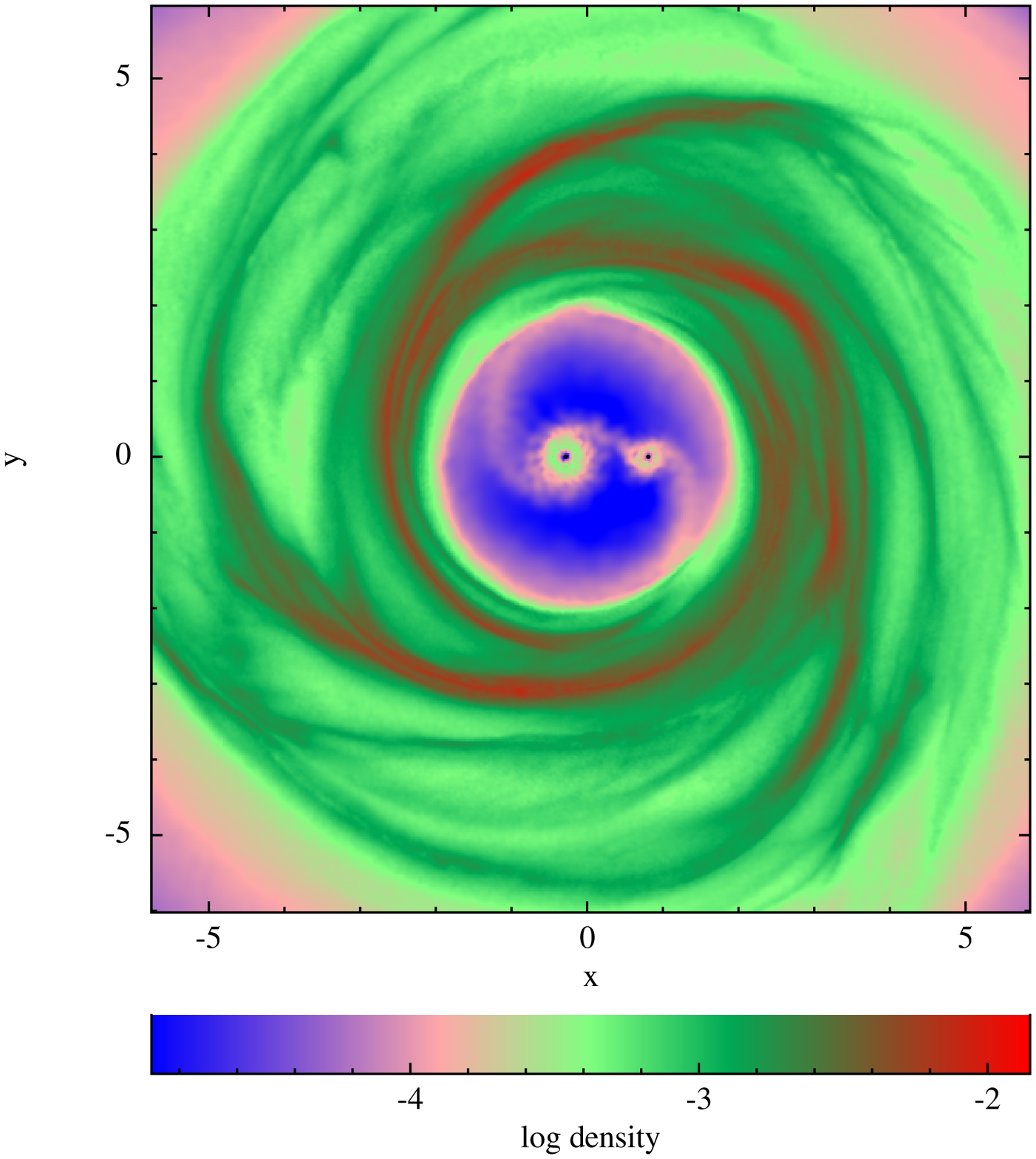}\\
\includegraphics[scale=0.3,clip=true,angle=0]{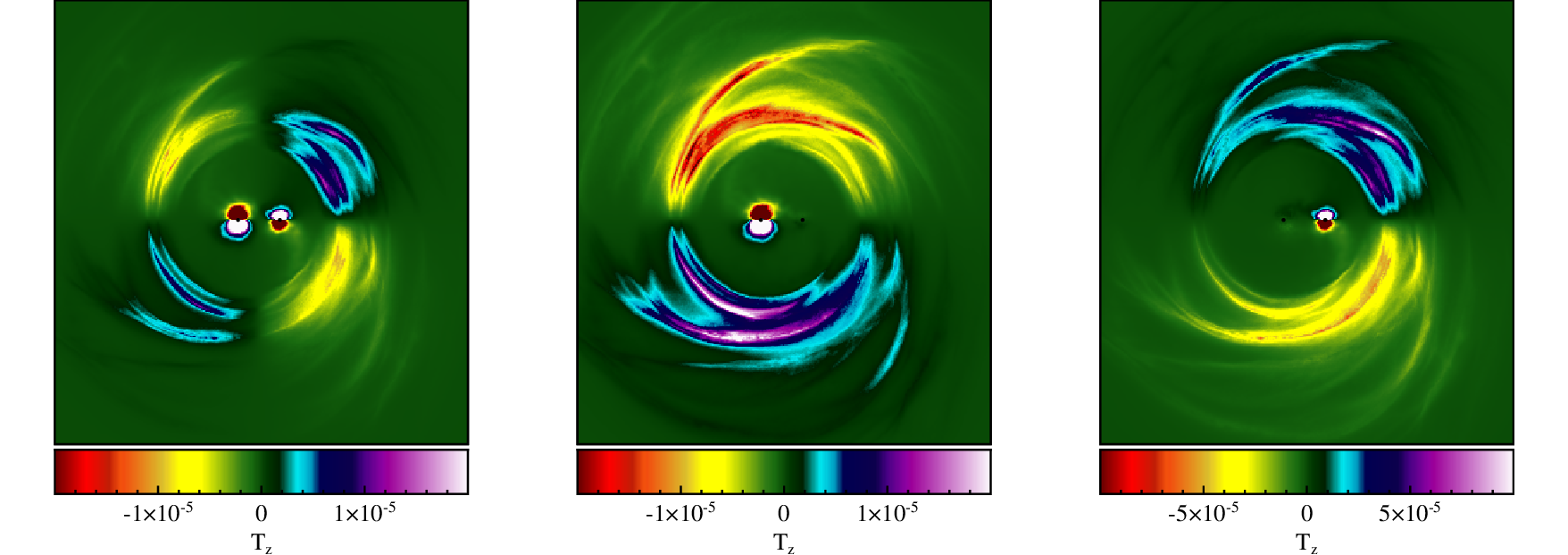}
\caption{Top panel: typical instantaneous logarithmic surface mass density of the
  \textit{iso05} simulation.  Bottom row: combined surface torque density
  exerted by the disc onto the binary (left panel), onto the primary
  BH (central panel) and onto the secondary BH (right panel).  All plots are in code units:
$G=M_0=a_0=1$.}
\label{disc_surfacedeniso}
\end{figure}

\begin{figure}
\centering
\includegraphics[scale=0.3,clip=true,angle=0]{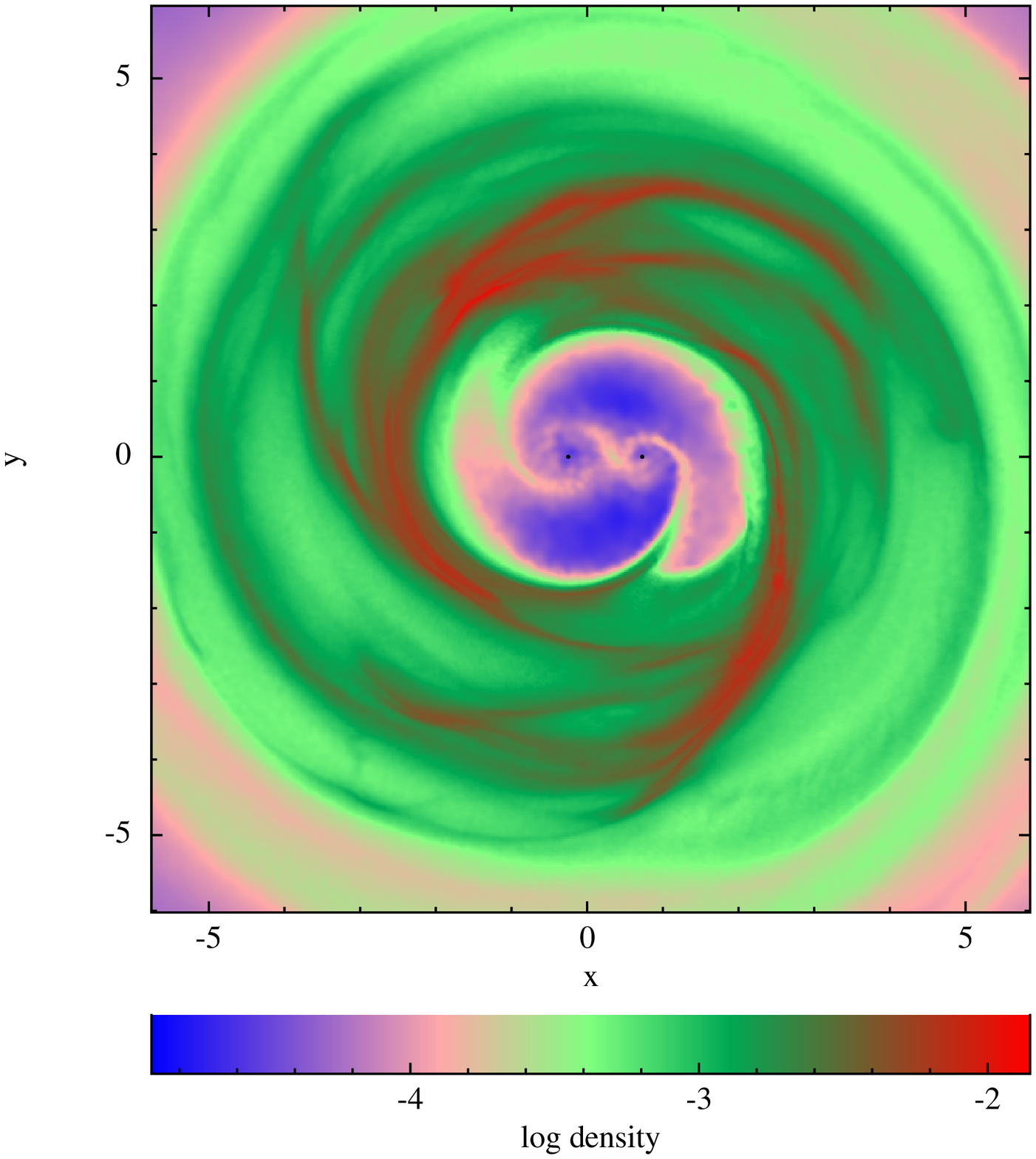}\\ 
\includegraphics[scale=0.3,clip=true,angle=0]{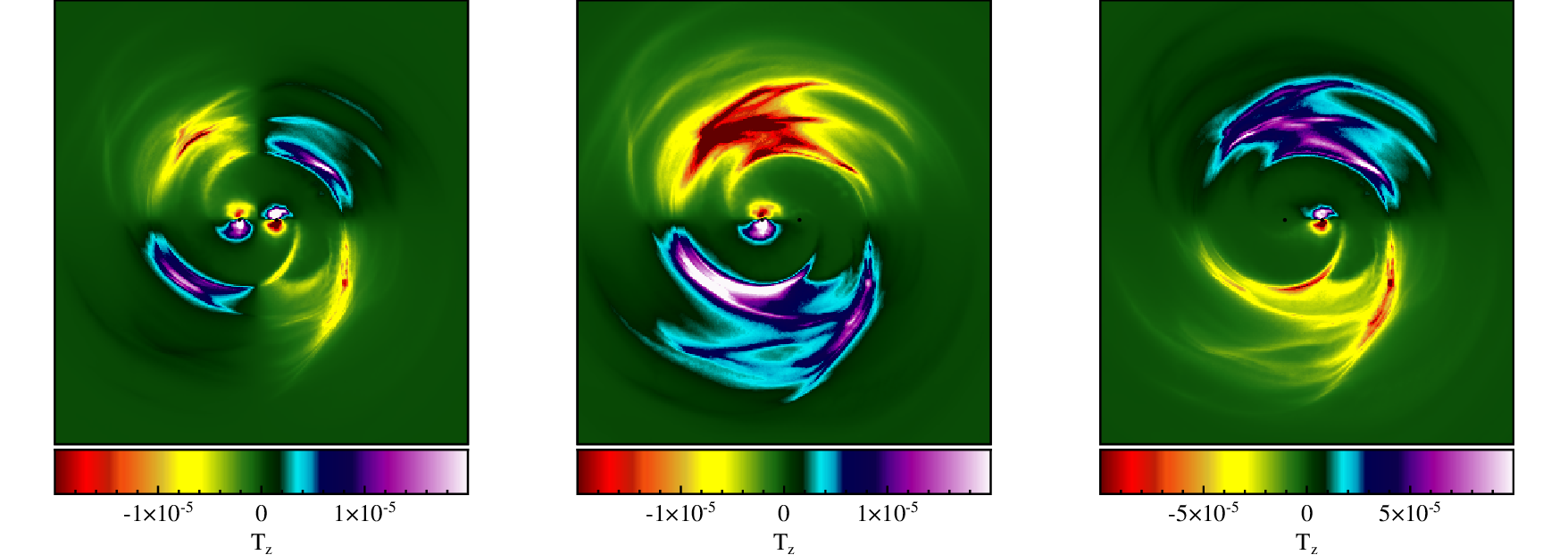}
\caption{Same as Fig.~\ref{disc_surfacedeniso} but for the \textit{adia05} run. Here the features
highlighted above are even more evident (especially the symmetric overdensities at 
the inner edge of the disc).}
\label{disc_surfacedenadia}
\end{figure}

It seems natural to compare the torque radial profiles obtained by our
simulations with linear theory perturbation, in which torque minima
and maxima are connected to outer Lindblad resonances (OLRs). It is in
fact tempting to associate the torque minimum at $r\approx1.6{a}$ with
the 2:1 OLR.  We should however be careful in pushing this
interpretation too far, since, as already pointed out by MM08 and 
S11, the assumptions of linear theory are not satisfied in this
context. Most importantly, looking at the upper panels of
Figs.~\ref{disc_surfacedeniso} and~\ref{disc_surfacedenadia}, we
notice that the region $r<2a$ is almost devoid of gas, and the streams
are almost radial. Mass fluxes reported in Tab.~\ref{tab:massflus}
clearly show that, at any radius, there are always fluxes of ingoing
and outgoing mass, resulting in a steady net inward flux consistent
with the accretion rate onto the two BHs. A strict OLR interpretation of the
torque minimum at $r\approx1.6a$ would instead require particles in circular orbit at
that radius, experiencing a secular effect due to the phase-coherent
periodic forcing of the binary; this is not what happens within the
cavity region.  The strongest 2:1 OLR is certainly responsible for the
evacuation of the gas close to the binary and for the formation and
maintenance of a cavity, however cannot be directly responsible for the coherent
torque seen in the cavity region. This is also supported by the fact that
MM08 and S11 find a minimum at the same location ($r\approx1.6-1.7a$)
for an equal mass binary, where the 2:1 resonance is absent (because
of the symmetry of the forcing potential), and the location of the
strongest OLR (3:2) would be at $r\approx1.3a$.  The strong negative
torque in the cavity region has a purely kinetic origin: material ripped
off the disc edge forms well defined streams {\it following} the two
BHs, which are clearly distinguishable in both the surface density
plots shown in Fig.~\ref{disc_surfacedeniso}
and~\ref{disc_surfacedenadia}. The streams are responsible for the
yellow tails following the two BHs at $\sim 1.5a$ in the torque density
panels, which lead to a net negative torque. Conversely, at
$r\gsim2a$, we have a well defined, almost circular disc, and the
torque density peaks at $r\approx2a$ and $r\approx2.5a$ can be
identified with the loci of the strong 3:1 and 4:1 OLRs
\citep{Artymowicz1994}.


\begin{figure}
\includegraphics[width=\linewidth]{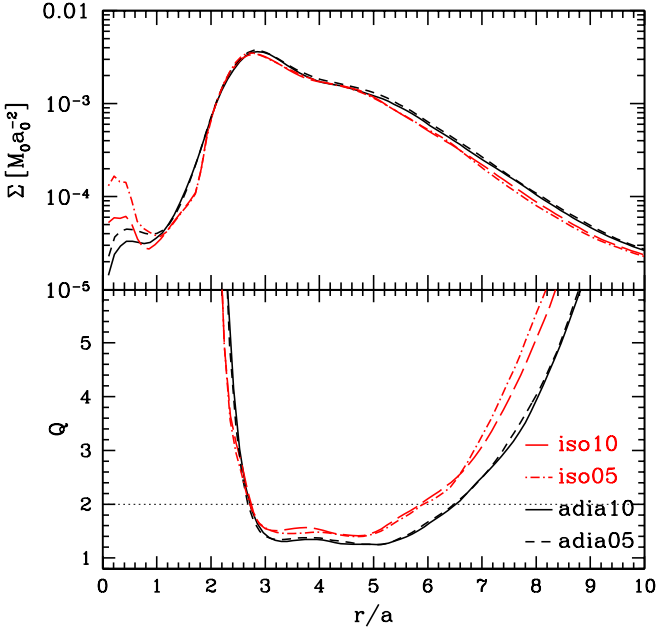}
\caption{Disc properties: average surface density (top panel),
and  Toomre parameter $Q$ (bottom panel) as a function of $r$. Line style denotes 
\textcolor{red}{iso10} (long dashed), \textcolor{red}{iso05} (dot dashed),
 \textcolor{black}{adia10} (solid) and \textcolor{black}{adia05} (dashed).
}
\label{disc}
\end{figure}

Our simulations also allow us to investigate the torques within the
binary corotation radius at $r<a$, a region often excised in
grid-based simulations (see MM08 and S11). Here we find strong
positive torques on both BHs. This is because the infalling material
approaches the BHs at super-Keplerian velocities, and bends in a
horseshoe fashion, exerting a net positive torque in front of them.
In fact, the maximum positive torque basically coincides with the
location of the two BHs (sharp peak at $r\approx 0.75{a}$ for the
secondary and a broader peak around $r\approx 0.3{a}$ for the primary,
see Fig.~\ref{torqueave}).  The very same effect, in the context of
planetary migration, was discussed by
\citet{LinPapa11}.\footnote{\citet{LinPapa11} found that such positive
  torque can in principle counter-balance the negative torque due to
  the disc and the tails, causing, in their case, angular momentum to
  be transferred {\it to} the planet, resulting in an {\it outwards}
  migration.} Note that the positive torque is related to this `stream
bending', and not directly to the small discs of gas orbiting each BH;
its magnitude is in fact similar in the {\it iso} and in the {\it
  adia} simulations, even though in the former, the mass in the
minidiscs is significantly larger as shown by the time and azimuthally
averaged surface density profiles in the upper panel of
Fig.~\ref{disc}.

\begin{table*}
  \caption{Accretion rates onto the binary and mass fluxes $F$ in units $M/P_0 \times 10^{-5}$ at two selected
distances to the binary CoM: $r_1=a$ and $r_2=1.5\,a$. $\dot{M}_1$ and $\dot{M}_2$ are the average accretion rates on
$M_1$ and $M_2$ respectively, while $\dot{M}$ is the sum of the two. At both selected distances, $F_{\rm in}$
and $F_{\rm out}$ represent the ingoing and outgoing fluxes, and $F_{\rm net}=F_{\rm in}-F_{\rm out}$ is the net
ingoing flux. All quantities have been averaged over $90$ binary orbits. 
\label{tab:massflus}}
\begin{center}
\begin{tabular}{cccccccccc}
\hline
\hline
& &&&& $r_1=a$  & & & $r_2=1.5a$ &\\
\hline
Model &$ \dot{M}_1 $  & $ \dot{M}_2$    &$ \dot{M} $   & $F_{\rm in_1}$ & $F_{\rm out_1}$ & $F_{\rm net_1}$ & $F_{\rm in_2}$ & $F_{\rm out_2}$ & $F_{\rm net_2}$\\
    \hline
$\mathtt{iso10}$   & $4.5$  & $10.7$  & $15.2 $ & $19.6$   & $4.5$  & $15.1  $ & $33.7 $& $18.8$ &$ 14.9$\\
$\mathtt{iso05}$   & $4.2$  & $9.2$  & $13.4 $   & $19.1 $  & $5.5 $  & $13.6$ & $31.7 $& $18.4 $ &$ 13.4$\\
   \hline
$\mathtt{adia10}$  &$3.8$  &$ 9.5 $& $13.3 $ & $17.6$  & $4.3$ & $13.3 $  &$ 42.8$& $29.7$& $13.2 $\\
$\mathtt{adia05}$  &$3.2 $ &$ 4.6$ & $7.8 $  & $13.5$  & $5.8$ & $7.7$ &$ 36.0$& $28.4$& $7.6$\\
\hline
    \hline
  \end{tabular}
\end{center}
\end{table*}

\subsubsection{Disc structure and characteristic torque frequencies}
\label{sec:discstructure}
The surface density profiles in Fig.~\ref{disc_surfacedeniso} and
~\ref{disc_surfacedenadia} highlight some clear differences between
the {\it iso} and the {\it adia} runs: (i) the appearance and shape of
the minidiscs; (ii) the amount of gas feeding the streams; (iii) the
definition of the inner disc edge.  In the \textit{iso} runs, we find
a large amount of gas forming mini-discs around both BHs and an
almost empty cavity except for two tenuous streams connecting the outer edge 
to the mini-discs.  The edge of the disc is quite circular and well
defined. Conversely, due to the gas inside the cavity being hotter in
the \textit{adia} runs, the streaming activity is more violent with
thick spirals that do not settle into well defined
mini-discs, but rather form a diluted, $\infty$-shaped cloud around
the binary.  The edge of the disc is strongly disturbed by the ripped
out streams and is usually not circular.  The different streaming
activity is also confirmed by numbers in Tab.~\ref{tab:massflus},
where we collect the average inwards and outwards mass fluxes at
$r=1.5a$ and $r=a$. Mass fluxes are generally larger in the
\textit{adia} case. Note that this does not necessarily result in
larger accretion, as outgoing fluxes are also larger in this
case. Most of the material in the streams, does not end up in an
accretion flow, but is accelerated back to the disc (a sort of
'slingshot') impacting on the inner edge, an effect already seen and
extensively described by MM08 and S11. The amount of impacting gas is
$50\%$ larger in the \textit{adia} case, possibly contributing to the
inner edge destabilization and to the larger perturbation in the disc
structure.  The appearance of spirals in the disc is related to the
behaviour of the Toomre $Q$ parameter, shown in the bottom panel of
Fig.~\ref{disc}, which quantifies the degree of self-gravity of a
disc.  Not surprisingly the disc develops a spiral pattern in the
region $3a\,\lsim\, r\, \lsim\, 5a$, where, in fact, we find that the
Toomre parameter reaches its minimum value $Q\approx1.5$
\citep{Jorge09}. The shape of the spiral arms varies much in time, and
their definition is different from simulation to simulation. However,
at any time we find spiral structures with $2$, $3$ or $4$ arms, which
remain confined between $3a < r< 5a$.

The disc structures  highlighted above are reflected in the torques depicted in the lower panels of Fig.~\ref{disc_surfacedeniso}
 and \ref{disc_surfacedenadia}. The total torque shows a
typical quadrupolar structure, which leads to rapid oscillations averaging to zero in the disc body. 
Conversely, as already discussed, in the inner cavity we can appreciate the 
yellow tails following the two BHs at$\sim 1.5a$, leading to a net negative torque. 

\begin{figure} 
\includegraphics[width=\linewidth]{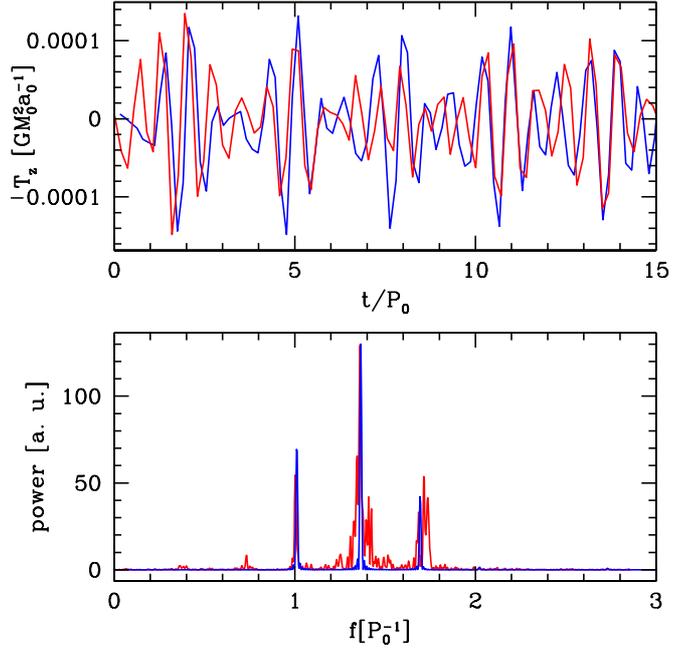}
\caption{Simple test model for the periodicity generated in the outside disc, i.e.
at $r>a$. In the top panel we show the temporal evolution of the torque, whereas in the
bottom panel we show the Fourier transform.
In each panel, the \textcolor{blue}{blue} curve is obtained by placing three point 
masses at distance $1.6a$, $2.1a$, $3.5a$ from a circular binary, and the
one \textcolor{red}{red} is the torque found in the {\it iso10} simulation. 
}
\label{semianaltoy}
\end{figure}

The main peaks observed in the torque power spectra must therefore arise from the structures we just
described. In particular we identified the overdensities at $r\approx 2a$ in the inner rim of the disc, and
the spiral structures at $r\gsim 3a$ in the main body of the disc. The main torque frequencies must come 
from the interaction between the forcing binary quadrupolar potential and such overdensities propagating in the 
disc. To test this hypothesis, we mimic the situation by placing test particles in circular orbits at $r=2a$ and 
$r=3.5a$ around a circular binary. We compute the torques and show them as \textcolor{blue}{blue} lines 
in Fig.~\ref{semianaltoy}.  
The natural frequency between a quadrupolar potential oscillating with frequency $f_1$ and an overdensity
orbiting at frequency $f_2$ is the beat frequency $2f_1-f_2$, and this is what we see in the Fourier
spectrum. The beats between the binary and the particles at $r=2a$ and at $r=3.5a$ give rise to sharp peaks seen at 
$f\approx 1.35P_0^{-1}$ and $f\approx 1.7P_0^{-1}$, respectively. The fact that such peaks are much sharper in the {\it iso} 
simulations stems from two facts: (i) the binary does not significantly shrink in the {\it iso} runs, limiting
the broadening of the spectral feature and (ii) the disc features (disc edge and spiral arms) are much more 
defined in this case, and the torques are well localized. This can be also seen by comparing the much neater
torque structure in the bottom panel of Fig.~\ref{disc_surfacedeniso} to the blurry one  of
 Fig.~\ref{disc_surfacedenadia}. 
The $f\approx P_0^{-1}$ peak of the \textcolor{blue}{blue} line in Fig.~\ref{semianaltoy} is obtained by 
placing a third particle at $r=1.6a$, a separation
corresponding to the maximum negative torque inside the cavity. However, Fig.~\ref{torquetotnew} shows that the peak at 
$f\approx P_0^{-1}$ is stronger at $r>1.8a$ than in the $a<r<1.8a$ region, meaning that it must be mostly directly 
related to the periodicity at which the binary rips gas off the disc edge (i.e. the binary period), and not to 
a specific radius in the cavity. Finally, as already noticed, torques in the binary region are related to the 
mass fuelling the two BHs, and therefore show periodicities related to the binary ($f=P_0^{-1}$), to the 
inner rim of the disc ($f\approx0.25P_0^{-1}$) and to the beat between the two ($f\approx 0.75P_0^{-1}$). Note that the latter 
two are much more evident in the {\it adia} runs, where the disc rim overdensities feeding the accretion streams
are more pronounced.

\section{Accretion}
\label{sec:spin}

In Tab.~\ref{tab:massflus} we also show the average mass accretion
rates. We recall here that a particle is considered accreted when it
crosses the sink radius of one of the BHs, 
In this sense, $r_{\rm sink}$ defines  a small
excision region around each BH. However,  
almost all the accreted particles are bound and lie well within the Roche 
lobe of the accreting BH.
The mass accretion rate on both BHs is almost independent on the 
sink radius in the {\it iso} simulations, where instreaming gas settles 
in well defined accretion discs progressively loosing 
angular momentum. In the {\it adia} simulations, this is also true for $M_1$, but 
the accretion rate 
onto $M_2$ scales linearly with the 
sink radius, suggesting a direct relation to the BH cross section 
(defined by the sink radius itself).
As a consequence,
we are likely overestimating accretion on this hole for the 
${\it adia}$ runs.
We should state here, that we are not interested 
in whether the particle is accreted immediately or is dragged along in a minidisc:
for the purpose of linear momentum transfer from the disc to the BH, the
two  processes are equivalent.
The physics 
of the minidiscs, beyond resolution issues, is likely to be radiation 
dominated, and is certainly not well described in our simulations. 

It is particularly interesting to compare the accretion rates to the
mass flow rates. Firstly we notice that in all simulations the
accretion rate and the mass flow rates at $r=a$ and $r=1.5a$ are
nearly the same, meaning that the system is in a steady state
configuration. Interestingly, the mass accretion rate is $\sim 25\%$
lower than the inflow rate at $r=a$. This means that not all the
material crossing $r=a$ is accreted, an assumption commonly adopted in
grid simulations where the $r<a$ region in excised. Part of the gas
suffers a slingshot impacting back to the disc, affecting further the
disc--binary angular momentum transfer.

\begin{figure}
\includegraphics[width=\linewidth]{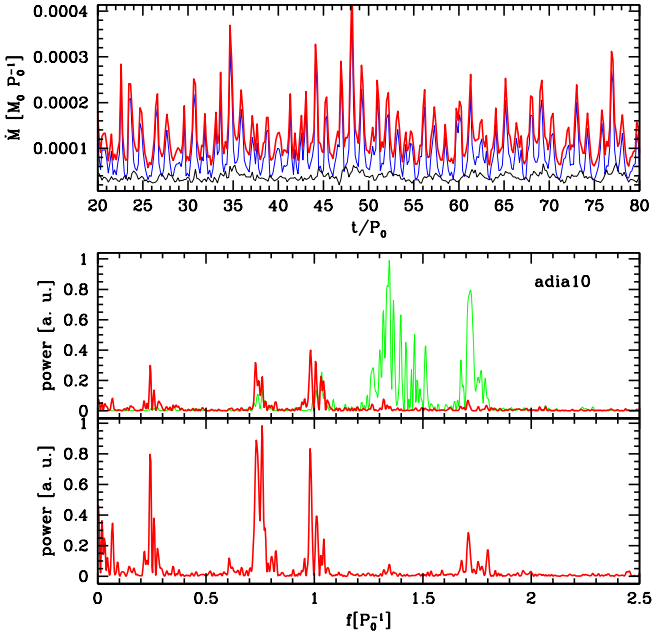}
\caption{Accretion onto the two BHs. Top box: accretion 
rate as a function of time. The \textcolor{red}{red} line is the total accretion rate while the 
\textcolor{blue}{blue} and \textcolor{black}{black} lines are the accretion rates onto the secondary and the primary
respectively. Lower box: Fourier transform of the total accretion rate (lower panel) and of the 
torques (upper panel). In the upper panel, the \textcolor{green}{green} line is the total torque 
(highest peak normalized to $1$) and the \textcolor{red}{red} line is the torque exerted by the material at 
$r<a$ (multiplied by ten). 
}
\label{accfou}
\end{figure}

In the top panel of Fig.~\ref{accfou} we show the temporal evolution
of the accretion rate on the primary hole, $\dot{M}_1$, on the
secondary hole, $\dot{M}_2$, and the sum of the two, $\dot{M}$. As
already found in \cite{Jorge09, Roedig2011}, $\dot{M}_2$ is a factor $\sim 2$
larger than $\dot{M}_1$, due to the closer interaction between the
secondary BH and the disc. The relative mass growth $\delta{M}/M$ is
therefore about six time faster for the secondary BH, implying the
binary mass ratio will tend toward $q\approx 1$ if this condition
persists over the entire evolution of the system.  Accretion rates are
strongly modulated in time, and notably, $\dot{M}_2$ shows a striking
periodicity related to the binary orbital period, whereas $\dot{M}_1$
is dominated by longer timescale fluctuation, related to the
overdensities developing at the inner rim of the circumbinary disc.

The Fourier transform of $\dot{M}$ is given in the lower panel of
Fig.~\ref{accfou}. We observe the usual three distinctive peaks
observed in the torques coming from the binary region, as shown by the
central panel of the figure.  It is clear that the accretion and
inner-torque periodicities are intimately connected, both reflecting
the temporal evolution of the mass inflow in the binary region (ii).

Finally we can check the orientation of the angular momentum vector of
the accreted material in the reference frame of the accreting BH. This
number quantifies the level of `coherence' of the accretion flow, and
has important consequences for the individual BH spin evolution and
the magnitude of gravitational recoil at the coalescence 
\citep{Bogdanovic:2007hp,Dotti:2009vz,Kesden10,Volonteri10,Lousto11,Lousto12}. At each snapshot we therefore compute the average $L_z$ and
${\bf L}$ of the accreted particles {\it in the accreting BH reference
  frame}, and the angle $\theta=\arccos{(L_z/{\bf L}})$ defining
the degree of misalignment with respect to the $z$ axis.  
Our results are similar to what already found by \citet{Dotti2009b},
although in the cited study the spin evolution was studied only before
the gas scouring. 
More specifically, an isothermal EoS leads to extremely coherent
accretion flows, with fluctuations in the angular momentum direction
of the mini-discs of few degrees. In the adiabatic case, the
orientation of the minidisc orbiting the secondary changes by at most
$\approx 20$ degrees, and has an average excursion of 
7 degrees, 50\% bigger than the oscillations in the primary
disc. Assuming that the BH spins have efficiently aligned with the
circumbinary disc before the opening of the cavity \citep{Dotti:2009vz}
this implies that the efficient spin-up of the binary will continue
following cavity opening.

\section{Comparison with previous works}
\label{sec:comparison}

The binary evolution and torque structure have been previously
investigated by MM08 and S11, as mentioned in the introduction. In
particular, we can compare results regarding the average gravitational
torques and the mass accretion rates. While both of them can be
expressed in several ways, we find it practical to normalize these
quantities to the disc mass, $M_d$, initial binary period, $P_0$, and 
initial binary angular momentum magnitude, $L_0$. In these units we can write:
\begin{equation}
\langle T_{[1,\infty]}\rangle=\Gamma_1 L_0 M_d P_0^{-1}
\end{equation}
\begin{equation}
\dot{M}=\Gamma_2 M_dP_0^{-1}
\end{equation}
Here $M_d=\pi \Sigma_{\rm p} r_{\rm p}^2$ is the disc mass computed
using the peak surface density\footnote{Note that we have a physical
  disc mass in our simulations, but we use this quantity to ease the
  comparison with MM08 and S11.} $\Sigma_{\rm p}$. In both MM08 and
S11, $r_{\rm p}\approx 3a$, close to what we find in our discs (see
Fig.~\ref{disc}). Before proceeding with this comparison we should
mention several issues that has to be considered. Firstly
$L_0=(M_0/4)(GM_0a_0)^{1/2}$ for the circular equal mass binary simulated by
MM08 and S11, whereas in our simulation with $q=1/3$ we have
$L_0=(3M_0/16)(GM_0a_0)^{1/2}$ (assuming a circular binary). Secondly, 
in MM08 and S11, the
concept of `accretion rate' is defined by measuring the mass flux through
the inner boundary, placed at $r=a$ for MM08 and $r=0.8a$ for S11. As
already discussed (see Tab.~\ref{tab:massflus}) $\approx25\%$ of the
mass crossing $r=a$ is accelerated back to the disc edge extracting
angular momentum from the binary (an effect already seen and described
extensively by MM08 and S11 for the material inflowing in the cavity but
not reaching the excised region). We can therefore consider MM08 and
S11 to overestimate the accretion rate by a factor of $25\%$. Lastly,
MM08 and S11 can compute gravitational torques only for $r \gtrsim a$,
outside the binary corotation radius\footnote{S11 already pointed out
  the change of sign of the torques at $r=a$.}. As shown by both
studies (and independently recovered by us), gravitational torques
{\it exerted by the disc elements on the binary} are negative in this
region.

For the sake of the comparison we use the {\it iso05} and the {\it
  adia05} runs. In the torque computation, we find
$\Gamma_1=-3.5\times 10^{-3}$ for the former and $\Gamma_1=-5.3\times
10^{-3}$ for the latter. This can be compared to $-1.26\times10^{-3}$
of MM08 and $-1.8\times10^{-2}$ of S11. Our gravity torques are a
factor 3--4 larger than MM08 and a factor 4--5 smaller than
S11. Concerning the accretion, we get $\Gamma_2=1.1\times 10^{-3}$ for
the {\it iso05} run and $\Gamma_2=6\times 10^{-4}$ for the {\it
  adia05} run, to be compared to $\Gamma_2=1.2\times 10^{-4}$ of MM08
and $\Gamma_2=6\times 10^{-3}$ for S11.  Our accretion rates are a
factor 5--10 larger than MM08 and a factor 5--10 smaller than S11.  The
similar scaling between $\langle T_{[1,\infty]}\rangle$ and $\dot{M}$
is not surprising, since, as discussed in Section \ref{sec:torx}, the
former is directly linked to the streams fuelling the BHs.

\section{Discussion and conclusions}
\label{sec:conclusion}
The evolution of BHBs in circumbinary discs remains largely an open
issue. Here we carried out a detailed study of the
torques acting on the BHB, including the effect of the outer disc, the
mass flowing in the cavity and, for the first time, of the gas entering
the BHB region and eventually accreted onto the two BHs. We paid
particular attention to investigating the individual contribution of 
different disc regions and we aimed at separating the effect of 
accretion from the
effect of gravitational torques. The emerging picture is, in general, more
complex than the often adopted simple assumption that resonant torques
arising at the disc inner edge transfer angular momentum from the BHB
to the disc, causing the orbital decay \citep[e.g.][]{papaloizou77,GoldreichTremaine80,Ivanov99,Armitage:2002,Haiman2009}.

Firstly, as already pointed out by a number of authors (e.g. MM08;
\citet{Jorge09,Roedig2011};S11), the presence of a BHB clearing a cavity in the disc {\it does not}
prevent gas inflows and eventual accretion onto the two BHs. Such
inflowing gas has a non negligible role in defining the BHB--disc
mutual evolution; it forms streams that follow the BHs, eventually
exerting a net {\it negative} torque in the cavity region. The inflows
catch-up with the BHs at a super-Keplerian speed, bend in front of
them, thus exerting a net {\it positive} torque inside the binary
corotation radius (i.e. at $r<a$). We therefore conclude that the
origin of the dominating gravitational torques on the binary is purely
kinematic, and can be understood without directly invoking resonances 
of the forcing binary potential with the disc.

The strong positive gravitational torque is related to the gas inflows
only, and not to the formation of minidiscs around the two BHs nor to
the eventual accretion. This becomes clear by inspecting
Figs.~\ref{torqueave},~\ref{disc} and Tab.~\ref{tab:massflus};
although there is a factor up to ten difference in the mass bound to
the BHs in minidiscs, and a factor of three difference in the
accretion rate among the simulations, the positive torque is almost
the same. This is because the gravitational torque is only related to
the {\it inflowing} mass bending in front of the BHs, which is of the
same order in all simulations. Note that different thermodynamics lead
to different negative/positive torque balance. Although this might be
an artifact of the sudden change of EoS in the {\it iso} simulations,
such a balance has to be better understood in order to assess the
fate of the binary.

Torques exerted by gas in the main body of the disc (excluding the disc edge) are instead
negligible for the long term evolution of the system, since they
average to zero (Fig.~\ref{torqueave}).  Local torque
maxima can be associated to the 3:1 and 4:1 OLRs
\citep{Artymowicz1994}. Given the nature of the forcing potential, a
quadrupolar torque pattern naturally develops
(Figs.~\ref{disc_surfacedeniso} and \ref{disc_surfacedenadia}) causing
a time oscillating torque (Fig.~\ref{torquetotnew}) with
characteristic frequencies related to the beat between twice the
binary frequency and the characteristic orbital frequencies of
inhomogeneities developing in the disc in form of inner edge
overdensities and spiral arms (Fig.~\ref{semianaltoy}).

Accretion plays an important role in the binary angular momentum
budget. The binary can actually gain angular momentum even if it
shrinks (see also S11) and its eccentricity increases. This is because 
accretion deposits a significant amount of angular momentum in the system by
increasing its mass and especially its mass ratio (see Eq.~(\ref{eqlzdec})). 
The peculiar case of {\it iso05} is quite interesting: here the total
gravitational torque exerted by the gas is almost null
(Fig.~\ref{torqueave}).  However, the BHB--disc mutual torques are
responsible for the eccentricity growth, meaning that, if accretion
were neglected, the BHB would be forced to expand, as shown in the 
upper-right panel of Fig.~\ref{delta_ae}. In this particular case
case, accretion is responsible for the shrinkage. Although whether
this happens in Nature is questionable, it highlights the complexity
of the BHB--disc interplay.

We should, nonetheless, be careful in drawing any conclusion about
accretion from our simulations. If we scale our system to the fiducial
binary of \citet{Roedig2011} and set $M_1=2.6\times 10^6\msun$ and
$a_0=\, 0.039 \,$ pc, then the accretion rate we find corresponds to
about 20--40$\dot{M}_{\rm Edd}$ for the secondary BH, and
4--7$\dot{M}_{\rm Edd}$ for the primary BH. In such situation, it is
likely that most of the gas which is {\it numerically} accreted in our
simulations will be expelled through outflows and winds.
In this case, if the gas binds to the BHs before being 
expelled, it transfers its linear momentum to the 
holes, even if not accreted \citep{nixon11a}. However, its mass
does not add-up to the binary, making the question of angular 
momentum transfer more delicate and worthy of deeper and more refined 
investigations.

We numerically investigated different EoSs and sink radii $r_{\rm sink}$. The
numerical size of $r_{\rm sink}$ has a minor impact on the general
behaviour of the system, affecting the mass bound to the BH in
minidiscs and the accretion rate in the {\it adia} case. Conversely,
the adopted EoS has a major impact on the results.  In the {\it iso}
runs, tenuous streams leak from an almost circular disc edge and fuel two
well defined mini-discs, whereas streaming activity is more violent in
the {\it adia} case, with thick spirals leaking from a deeply
distorted disc edge, forming a diluted, $\infty$-shaped cloud around
the binary.  This difference in streaming activity affects both the
overall structure of the outer disc and the long-term binary evolution
(as mentioned above). Although the extrapolation of the results of all
our four simulations would imply a fast BHB coalescence (in less than
$10^7$ yr, for the fiducial binary parameters considered above), the
almost perfect cancellation of the net gravitational torque in the
{\it iso} simulations suggest that more realistic models for gas
thermodynamics will be necessary to make clear-cut statements on this
issue.

\section*{Acknowledgements}
We thank the anonymous referee for suggesting necessary clarifying
comments.
We are indebted to Monica Colpi for lively and inspiring discussions.  
CR is grateful to Nico Budewitz for HPC support.  
The computations were performed on the {\it datura}
cluster of the AEI. JC acknowledges support from FONDAP (15010003),
FONDECYT (11100240), Basal (PFB0609), and VRI-PUC (Inicio 16/2010).
MD and JC appreciate the warm hospitality at the AEI. 
This work has, in part, been supported by the Transregio 7 "Gravitational Wave
Astronomy" financed by the Deutsche Forschungsgemeinschaft DFG (German
Research Foundation).

\bibliographystyle{aa}
\bibliography{aeireferences.bib}

\begin{thebibliography}{63}
\expandafter\ifx\csname natexlab\endcsname\relax\def\natexlab#1{#1}\fi

\bibitem[{{Armitage} \& {Natarajan}(2002)}]{Armitage:2002}
{Armitage}, P.~J. \& {Natarajan}, P. 2002, Astrophys. J. Lett., 567, L9

\bibitem[{{Armitage} \& {Natarajan}(2005)}]{Armitage:2005}
{Armitage}, P.~J. \& {Natarajan}, P. 2005, \apj, 634, 921

\bibitem[{{Artymowicz} \& {Lubow}(1994)}]{Artymowicz1994}
{Artymowicz}, P. \& {Lubow}, S.~H. 1994, Astrophys. J., 421, 651

\bibitem[{{Bate} \& {Bonnell}(1997)}]{Bate97}
{Bate}, M.~R. \& {Bonnell}, I.~A. 1997, Mon. Not. R. Astron. Soc., 285, 33

\bibitem[{{Bate} {et~al.}(1995){Bate}, {Bonnell}, \& {Price}}]{Bate95}
{Bate}, M.~R., {Bonnell}, I.~A., \& {Price}, N.~M. 1995, \mnras, 277, 362

\bibitem[{{Begelman} {et~al.}(1980){Begelman}, {Blandford}, \&
  {Rees}}]{begelman80}
{Begelman}, M.~C., {Blandford}, R.~D., \& {Rees}, M.~J. 1980, Nature, 287, 307

\bibitem[{{Bogdanovi{\'c}} {et~al.}(2007){Bogdanovi{\'c}}, {Reynolds}, \&
  {Miller}}]{Bogdanovic:2007hp}
{Bogdanovi{\'c}}, T., {Reynolds}, C.~S., \& {Miller}, M.~C. 2007, ApJ, 661,
  L147

\bibitem[{{Bryden} {et~al.}(1999){Bryden}, {Chen}, {Lin}, {Nelson}, \&
  {Papaloizou}}]{Bryden99}
{Bryden}, G., {Chen}, X., {Lin}, D.~N.~C., {Nelson}, R.~P., \& {Papaloizou},
  J.~C.~B. 1999, Astrophys. J., 514, 344

\bibitem[{{Cuadra} {et~al.}(2009){Cuadra}, {Armitage}, {Alexander}, \&
  {Begelman}}]{Jorge09}
{Cuadra}, J., {Armitage}, P.~J., {Alexander}, R.~D., \& {Begelman}, M.~C. 2009,
  \mnras, 393, 1423

\bibitem[{{Cuadra} {et~al.}(2006){Cuadra}, {Nayakshin}, {Springel}, \& {Di
  Matteo}}]{Cuadra2006a}
{Cuadra}, J., {Nayakshin}, S., {Springel}, V., \& {Di Matteo}, T. 2006, Mon.
  Not. R. Astron. Soc., 366, 358

\bibitem[{{Davies} {et~al.}(2004{\natexlab{a}}){Davies}, {Tacconi}, \&
  {Genzel}}]{Davies04a}
{Davies}, R.~I., {Tacconi}, L.~J., \& {Genzel}, R. 2004{\natexlab{a}}, \apj,
  613, 781

\bibitem[{{Davies} {et~al.}(2004{\natexlab{b}}){Davies}, {Tacconi}, \&
  {Genzel}}]{Davies04b}
{Davies}, R.~I., {Tacconi}, L.~J., \& {Genzel}, R. 2004{\natexlab{b}}, \apj,
  602, 148

\bibitem[{{Dotti} {et~al.}(2007){Dotti}, {Colpi}, {Haardt}, \&
  {Mayer}}]{Dotti07}
{Dotti}, M., {Colpi}, M., {Haardt}, F., \& {Mayer}, L. 2007, \mnras, 379, 956

\bibitem[{{Dotti} {et~al.}(2009){Dotti}, {Ruszkowski}, {Paredi}, {Colpi},
  {Volonteri}, \& {Haardt}}]{Dotti2009b}
{Dotti}, M., {Ruszkowski}, M., {Paredi}, L., {et~al.} 2009, \mnras, 396, 1640

\bibitem[{{Dotti} {et~al.}(2012){Dotti}, {Sesana}, \& {Decarli}}]{Dotti12}
{Dotti}, M., {Sesana}, A., \& {Decarli}, R. 2012, Advances in Astronomy, 2012

\bibitem[{{Dotti} {et~al.}(2010){Dotti}, {Volonteri}, {Perego}, {Colpi},
  {Ruszkowski}, \& {Haardt}}]{Dotti:2009vz}
{Dotti}, M., {Volonteri}, M., {Perego}, A., {et~al.} 2010, Mon. Not. R. Astron.
  Soc., 402, 682

\bibitem[{{Downes} \& {Solomon}(1998)}]{Downes98}
{Downes}, D. \& {Solomon}, P.~M. 1998, \apj, 507, 615

\bibitem[{{Dutrey} {et~al.}(1994){Dutrey}, {Guilloteau}, \& {Simon}}]{Dutrey94}
{Dutrey}, A., {Guilloteau}, S., \& {Simon}, M. 1994, Astronomy \& Astrophysics,
  286, 149

\bibitem[{{Escala} {et~al.}(2005){Escala}, {Larson}, {Coppi}, \&
  {Mardones}}]{Escala2005}
{Escala}, A., {Larson}, R.~B., {Coppi}, P.~S., \& {Mardones}, D. 2005, \apj,
  630, 152

\bibitem[{{Fabbiano} {et~al.}(2011){Fabbiano}, {Wang}, {Elvis}, \&
  {Risaliti}}]{Fabbiano11}
{Fabbiano}, G., {Wang}, J., {Elvis}, M., \& {Risaliti}, G. 2011, Nature, 477,
  431

\bibitem[{{Foreman} {et~al.}(2009){Foreman}, {Volonteri}, \&
  {Dotti}}]{Foreman09}
{Foreman}, G., {Volonteri}, M., \& {Dotti}, M. 2009, \apj, 693, 1554

\bibitem[{{Gammie}(2001)}]{Gammie01}
{Gammie}, C.~F. 2001, Astrophys. J., 553, 174

\bibitem[{{Goldreich} \& {Tremaine}(1980)}]{GoldreichTremaine80}
{Goldreich}, P. \& {Tremaine}, S. 1980, \apj, 241, 425

\bibitem[{{Greve} {et~al.}(2009){Greve}, {Papadopoulos}, {Gao}, \&
  {Radford}}]{Greve09}
{Greve}, T.~R., {Papadopoulos}, P.~P., {Gao}, Y., \& {Radford}, S.~J.~E. 2009,
  \apj, 692, 1432

\bibitem[{{G{\"u}ltekin} {et~al.}(2009){G{\"u}ltekin}, {Richstone}, {Gebhardt},
  {Lauer}, {Tremaine}, {Aller}, {Bender}, {Dressler}, {Faber}, {Filippenko},
  {Green}, {Ho}, {Kormendy}, {Magorrian}, {Pinkney}, \& {Siopis}}]{gultekin09}
{G{\"u}ltekin}, K., {Richstone}, D.~O., {Gebhardt}, K., {et~al.} 2009, \apj,
  698, 198

\bibitem[{{G{\"u}nther} \& {Kley}(2002)}]{Gunther02}
{G{\"u}nther}, R. \& {Kley}, W. 2002, Astron. \& Astrophys., 387, 550

\bibitem[{{G{\"u}nther} {et~al.}(2004){G{\"u}nther}, {Sch{\"a}fer}, \&
  {Kley}}]{Gunther04}
{G{\"u}nther}, R., {Sch{\"a}fer}, C., \& {Kley}, W. 2004, Astron. \&
  Astrophys., 423, 559

\bibitem[{{Haiman} {et~al.}(2009){Haiman}, {Kocsis}, \& {Menou}}]{Haiman2009}
{Haiman}, Z., {Kocsis}, B., \& {Menou}, K. 2009, Astrophysical Journal, 700,
  1952

\bibitem[{{Hayasaki}(2009)}]{hayasaki09}
{Hayasaki}, K. 2009, Publications of the Astronomical Society of Japan, 61, 65

\bibitem[{{Hayasaki} {et~al.}(2008){Hayasaki}, {Mineshige}, \& {Ho}}]{haya08}
{Hayasaki}, K., {Mineshige}, S., \& {Ho}, L.~C. 2008, Astrophys. J., 682, 1134

\bibitem[{{Hayasaki} {et~al.}(2007){Hayasaki}, {Mineshige}, \&
  {Sudou}}]{hayasaki07}
{Hayasaki}, K., {Mineshige}, S., \& {Sudou}, H. 2007, Publications of the
  Astronomical Society of Japan, 59, 427

\bibitem[{{Hennawi} {et~al.}(2006){Hennawi}, {Strauss}, {Oguri}, {Inada},
  {Richards}, {Pindor}, {Schneider}, {Becker}, {Gregg}, {Hall}, {Johnston},
  {Fan}, {Burles}, {Schlegel}, {Gunn}, {Lupton}, {Bahcall}, {Brunner}, \&
  {Brinkmann}}]{Hennawi06}
{Hennawi}, J.~F., {Strauss}, M.~A., {Oguri}, M., {et~al.} 2006, Astronomical
  Journal, 131, 1

\bibitem[{{Ivanov} {et~al.}(1999){Ivanov}, {Papaloizou}, \&
  {Polnarev}}]{Ivanov99}
{Ivanov}, P.~B., {Papaloizou}, J.~C.~B., \& {Polnarev}, A.~G. 1999, Mon. Not.
  R. Astron. Soc., 307, 79

\bibitem[{{Kesden} {et~al.}(2010){Kesden}, {Sperhake}, \& {Berti}}]{Kesden10}
{Kesden}, M., {Sperhake}, U., \& {Berti}, E. 2010, \apj, 715, 1006

\bibitem[{{Khan} {et~al.}(2011){Khan}, {Just}, \& {Merritt}}]{Khan11}
{Khan}, F.~M., {Just}, A., \& {Merritt}, D. 2011, Astrophys. J., 732, 89

\bibitem[{Komossa {et~al.}(2003)Komossa, Burwitz, Hasinger, Predehl, Kaastra,
  \& Ikebe}]{Komossa:2002tn}
Komossa, S., Burwitz, V., Hasinger, G., {et~al.} 2003, Astrophys. J., 582, L15

\bibitem[{{Lin} \& {Papaloizou}(1986)}]{Lin86}
{Lin}, D.~N.~C. \& {Papaloizou}, J. 1986, Astrophys. J., 309, 846

\bibitem[{{Lin} \& {Papaloizou}(2012)}]{LinPapa11}
{Lin}, M.-K. \& {Papaloizou}, J.~C.~B. 2012, \mnras, 421, 780

\bibitem[{{Lodato} {et~al.}(2009){Lodato}, {Nayakshin}, {King}, \&
  {Pringle}}]{Lodato2009}
{Lodato}, G., {Nayakshin}, S., {King}, A.~R., \& {Pringle}, J.~E. 2009, Mon.
  Not. R. Astron. Soc., 398, 1392

\bibitem[{{Lousto} \& {Zlochower}(2011)}]{Lousto11}
{Lousto}, C.~O. \& {Zlochower}, Y. 2011, Physical Review Letters, 107, 231102

\bibitem[{{Lousto} {et~al.}(2012){Lousto}, {Zlochower}, {Dotti}, \&
  {Volonteri}}]{Lousto12}
{Lousto}, C.~O., {Zlochower}, Y., {Dotti}, M., \& {Volonteri}, M. 2012, \prd,
  85, 084015

\bibitem[{{Lubow} {et~al.}(1999){Lubow}, {Seibert}, \& {Artymowicz}}]{Lubow99}
{Lubow}, S.~H., {Seibert}, M., \& {Artymowicz}, P. 1999, Astrophys. J., 526,
  1001

\bibitem[{{MacFadyen} \& {Milosavljevi{\'c}}(2008)}]{MacFadyen2008}
{MacFadyen}, A.~I. \& {Milosavljevi{\'c}}, M. 2008, Astrophys. J., 672, 83

\bibitem[{{Mayer} {et~al.}(2007){Mayer}, {Kazantzidis}, {Madau}, {Colpi},
  {Quinn}, \& {Wadsley}}]{Mayer07}
{Mayer}, L., {Kazantzidis}, S., {Madau}, P., {et~al.} 2007, Science, 316, 1874

\bibitem[{{Mihos} \& {Hernquist}(1996)}]{mihos96}
{Mihos}, J.~C. \& {Hernquist}, L. 1996, \apj, 464, 641

\bibitem[{{Myers} {et~al.}(2007){Myers}, {Brunner}, {Richards}, {Nichol},
  {Schneider}, \& {Bahcall}}]{Myers07}
{Myers}, A.~D., {Brunner}, R.~J., {Richards}, G.~T., {et~al.} 2007, \apj, 658,
  99

\bibitem[{{Myers} {et~al.}(2008){Myers}, {Richards}, {Brunner}, {Schneider},
  {Strand}, {Hall}, {Blomquist}, \& {York}}]{Myers08}
{Myers}, A.~D., {Richards}, G.~T., {Brunner}, R.~J., {et~al.} 2008, \apj, 678,
  635

\bibitem[{{Nelson} {et~al.}(2000){Nelson}, {Papaloizou}, {Masset}, \&
  {Kley}}]{Nelson00}
{Nelson}, R.~P., {Papaloizou}, J.~C.~B., {Masset}, F., \& {Kley}, W. 2000, Mon.
  Not. R. Astron. Soc., 318, 18

\bibitem[{{Nixon} {et~al.}(2011){Nixon}, {Cossins}, {King}, \&
  {Pringle}}]{nixon11a}
{Nixon}, C.~J., {Cossins}, P.~J., {King}, A.~R., \& {Pringle}, J.~E. 2011, Mon.
  Not. R. Astron. Soc., 412, 1591

\bibitem[{{Papaloizou} \& {Pringle}(1977)}]{papaloizou77}
{Papaloizou}, J. \& {Pringle}, J.~E. 1977, Mon. Not. R. Astron. Soc., 181, 441

\bibitem[{{Preto} {et~al.}(2011){Preto}, {Berentzen}, {Berczik}, \&
  {Spurzem}}]{Preto11}
{Preto}, M., {Berentzen}, I., {Berczik}, P., \& {Spurzem}, R. 2011, Astrophys.
  J. Lett., 732, L26+

\bibitem[{{Price}(2007)}]{Price2007}
{Price}, D.~J. 2007, Publ. Astron. Soc. Aust., 24, 159

\bibitem[{{Rice} {et~al.}(2005){Rice}, {Lodato}, \& {Armitage}}]{Rice05}
{Rice}, W.~K.~M., {Lodato}, G., \& {Armitage}, P.~J. 2005, Mon. Not. R. Astron.
  Soc, 364, L56

\bibitem[{{Roedig} {et~al.}(2011){Roedig}, {Dotti}, {Sesana}, {Cuadra}, \&
  {Colpi}}]{Roedig2011}
{Roedig}, C., {Dotti}, M., {Sesana}, A., {Cuadra}, J., \& {Colpi}, M. 2011,
  Mon. Not. R. Astron. Soc, 415, 3033

\bibitem[{{Sanders} \& {Mirabel}(1996)}]{Sanders96}
{Sanders}, D.~B. \& {Mirabel}, I.~F. 1996, Annual Review of Astronomy and
  Astrophysics, 34, 749

\bibitem[{{Shakura} \& {Sunyaev}(1973)}]{ss73}
{Shakura}, N.~I. \& {Sunyaev}, R.~A. 1973, Astronomy \& Astrophysics, 24, 337

\bibitem[{{Shen} {et~al.}(2011){Shen}, {Richards}, {Strauss}, {Hall},
  {Schneider}, {Snedden}, {Bizyaev}, {Brewington}, {Malanushenko},
  {Malanushenko}, {Oravetz}, {Pan}, \& {Simmons}}]{Shen11}
{Shen}, Y., {Richards}, G.~T., {Strauss}, M.~A., {et~al.} 2011, Astrophys. J.
  Supplements, 194, 45

\bibitem[{{Shi} {et~al.}(2012){Shi}, {Krolik}, {Lubow}, \& {Hawley}}]{Shi11}
{Shi}, J.-M., {Krolik}, J.~H., {Lubow}, S.~H., \& {Hawley}, J.~F. 2012, \apj,
  749, 118

\bibitem[{{Springel}(2005)}]{Springel05}
{Springel}, V. 2005, \mnras, 364, 1105

\bibitem[{{Springel} {et~al.}(2005){Springel}, {Di Matteo}, \&
  {Hernquist}}]{Springel05b}
{Springel}, V., {Di Matteo}, T., \& {Hernquist}, L. 2005, \mnras, 361, 776

\bibitem[{{Volonteri} {et~al.}(2010){Volonteri}, {G{\"u}ltekin}, \&
  {Dotti}}]{Volonteri10}
{Volonteri}, M., {G{\"u}ltekin}, K., \& {Dotti}, M. 2010, \mnras, 404, 2143

\bibitem[{{Ward}(1997)}]{Ward97}
{Ward}, W.~R. 1997, Icarus, 126, 261

\bibitem[{{White} \& {Rees}(1978)}]{White78}
{White}, S.~D.~M. \& {Rees}, M.~J. 1978, Mon. Not. R. Astron. Soc., 183, 341

\end{thebibliography}

\end{document}